\documentclass[twocolumn, tighten]{aastex62}

\newcommand{\ntd}{N$_2$D$^+$}
\newcommand{\hnc}{HN$^{13}$C}
\newcommand{\cch}{c-C$_3$H$_2$}
\newcommand{\nth}{N$_2$H$^+$}
\newcommand{\hcn}{HC$_3$N}
\newcommand{\msun}{$M_{\sun}$}

\newcommand{\kms}{km s$^{-1}$}
\newcommand{\ta}{$T_{\rm A}^{\ast}$}

\newcommand{\pscore}{prestellar core}
\newcommand{\lscore}{late starless core}

\newenvironment{rotatepage}%
    {\pagebreak[4]\global\pdfpageattr\expandafter{\the\pdfpageattr/Rotate 90}}%
    {\pagebreak[4]\global\pdfpageattr\expandafter{\the\pdfpageattr/Rotate 0}}%

\usepackage{natbib,graphicx,morefloats,multirow,pbox,booktabs, makecell,amsmath}
\usepackage{hyperref}
\usepackage{soul}
\begin{document}

\title{Molecular Cloud Cores with High Deuterium Fraction: Nobeyama Single-Pointing Survey}

\correspondingauthor{Gwanjeong Kim}
\email{gj.kim@nao.ac.jp}

\author[0000-0003-2011-8172]{Gwanjeong Kim}
\affil{Nobeyama Radio Observatory, National Astronomical Observatory of Japan,
National Institutes of Natural Sciences,
Nobeyama, Minamimaki, Minamisaku, Nagano 384-1305, Japan}

\author[0000-0002-8149-8546]{Ken'ichi Tatematsu}
\affil{Nobeyama Radio Observatory, National Astronomical Observatory of Japan,
National Institutes of Natural Sciences,
Nobeyama, Minamimaki, Minamisaku, Nagano 384-1305, Japan}
\affiliation{Department of Astronomical Science,
SOKENDAI (The Graduate University for Advanced Studies),
2-21-1 Osawa, Mitaka, Tokyo 181-8588, Japan}

\author{Tie Liu}
\affiliation{Shanghai Astronomical Observatory, Chinese Academy of Sciences, 80 Nandan Road, Shanghai 200030, People’s Republic of China}

\author{Hee-weon Yi}
\affiliation{School of Space Research, Kyung Hee University, Seocheon-Dong, Giheung-Gu, Yongin-Si, Gyeonggi-Do, 446-701, South Korea}

\author[0000-0002-3938-4393]{Jinhua He}
\affiliation{Yunnan Observatories, Chinese Academy of Sciences, 396 Yangfangwang, Guandu District, Kunming, 650216, P. R. China}
\affiliation{Chinese Academy of Sciences South America Center for Astronomy, National Astronomical Observatories, CAS, Beijing 100101, China}
\affiliation{Departamento de Astronom\'ia, Universidad de Chile, Casilla 36-D, Santiago, Chile}

\author{Naomi Hirano}
\affiliation{Academia Sinica Institute of Astronomy and Astrophysics, 11F of Astronomy-Mathematics Building, AS/NTU. No.1, Sec. 4, Roosevelt Rd, Taipei 10617, Taiwan, R.O.C.}

\author{Sheng-Yuan Liu}
\affiliation{Academia Sinica Institute of Astronomy and Astrophysics, 11F of Astronomy-Mathematics Building, AS/NTU. No.1, Sec. 4, Roosevelt Rd, Taipei 10617, Taiwan, R.O.C.}

\author{Minho Choi}
\affiliation{Korea Astronomy and Space Science Institute,
Daedeokdaero 776, Yuseong, Daejeon 305-348, South
Korea}

\author[0000-0002-7125-7685]{Patricio Sanhueza}
\affiliation{National Astronomical Observatory of Japan,
National Institutes of Natural Sciences,
2-21-1 Osawa, Mitaka, Tokyo 181-8588, Japan}

\author[0000-0002-5310-4212]{L. Viktor T\'oth}
\affiliation{Department of Astronomy, E\"otv\"os Lor\'and Unviersity, P\'azm´any P\'eter s\'etny 1, 1117 Budapest, Hungary}

\author{Neal J. Evans II}
\affiliation{Department of Astronomy, The University of Texas at Austin, 2515 Speedway, Stop C1400, Austin, TX 78712-1205, USA}

\author{Siyi Feng}
\affiliation{Chinese Academy of Sciences Key Laboratory of FAST, National Astronomical Observatory of China, Datun Road 20, Chaoyang, Beijing, 100012, P. R. China}
\affiliation{National Astronomical Observatory of Japan, National Institutes of Natural Sciences, 2-21-1 Osawa, Mitaka, Tokyo 181-8588, Japan}
\affiliation{Academia Sinica Institute of Astronomy and Astrophysics, No.1, Sec. 4, Roosevelt Rd, Taipei 10617, Taiwan, Republic of China} 

\author{Mika Juvela}
\affiliation{Department of Physics, P.O.Box 64, FI-00014, University of Helsinki,
Finland}

\author{Kee-Tae Kim}
\affiliation{Korea Astronomy and Space Science Institute,
Daedeokdaero 776, Yuseong, Daejeon 305-348, South
Korea}

\author{Charlotte VASTEL}
\affiliation{IRAP, Université de Toulouse, CNRS, CNES, UPS, (Toulouse), France}

\author{Jeong-Eun Lee}
\affiliation{School of Space Research, Kyung Hee University, Seocheon-Dong, Giheung-Gu, Yongin-Si, Gyeonggi-Do, 446-701, South Korea}

\author{Quang Nguy$\tilde{\hat{e}}$n Lu'o'ng}
\affiliation{National Astronomical Observatory of Japan,
National Institutes of Natural Sciences,
2-21-1 Osawa, Mitaka, Tokyo 181-8588, Japan}
\affiliation{Korea Astronomy and Space Science Institute,
Daedeokdaero 776, Yuseong, Daejeon 305-348, South
Korea}
\affiliation{IBM, Canada}

\author{Miju Kang}
\affiliation{Korea Astronomy and Space Science Institute,
Daedeokdaero 776, Yuseong, Daejeon 305-348, South
Korea}

\author{Isabelle Ristorcelli}
\affiliation{RAP, CNRS (UMR5277), Universit´e Paul Sabatier, 9 avenue du Colonel Roche, BP 44346, 31028, Toulouse Cedex 4, France}

\author{O. Feh\'{e}r}
\affiliation{Institut de Radioastronomie Millimetrique, 300 Rue de la Piscine, 38406, Saint Martin d'Heres, France}

\author{Yuefang Wu}
\affiliation{Department of Astronomy, Peking University, 100871, Beijing, China}

\author{Satoshi Ohashi}
\affiliation{The Institute of Physical and Chemical Research (RIKEN), 2-1, Hirosawa, Wako-shi, Saitama 351-0198, Japan}

\author[0000-0002-7237-3856]{Ke Wang}
\affiliation{Kavli Institute for Astronomy and Astrophysics, Peking University, 5 Yiheyuan Road, Haidian District, Beijing 100871, China}

\author{Ryo Kandori}
\affiliation{National Astronomical Observatory of Japan,
National Institutes of Natural Sciences,
2-21-1 Osawa, Mitaka, Tokyo 181-8588, Japan}

\author{Tomoya Hirota}
\affiliation{National Astronomical Observatory of Japan,
National Institutes of Natural Sciences,
2-21-1 Osawa, Mitaka, Tokyo 181-8588, Japan}
\affiliation{Department of Astronomical Science,
SOKENDAI (The Graduate University for Advanced Studies),
2-21-1 Osawa, Mitaka, Tokyo 181-8588, Japan}

\author{Takeshi Sakai}
\affiliation{Graduate School of Informatics and Engineering, The University of Electro-Communications, Chofu, Tokyo 182-8585, Japan}

\author{Xing Lu}
\affiliation{National Astronomical Observatory of Japan,
National Institutes of Natural Sciences,
2-21-1 Osawa, Mitaka, Tokyo 181-8588, Japan}

\author{Mark A. Thompson}
\affiliation{Centre for Astrophysics Research, Science \& Technology Research Institute, University of Hertfordshire, Hatfield, AL10 9AB, UK}

\author{Gary A. Fuller}
\affiliation{Jodrell Bank Centre for Astrophysics, School of Physics and Astronomy, University of Manchester, Oxford Road, Manchester, M13 9PL, UK}

\author{Di Li}
\affiliation{National Astronomical Observatories, Chinese Academy of Sciences, Beijing, 100012, China}

\author{Hiroko Shinnaga}
\affiliation{Department of Physics, Kagoshima University, 1-21-35, Korimoto, Kagoshima, 890-0065, Japan}

\author{Jungha Kim}
\affiliation{National Astronomical Observatory of Japan,
National Institutes of Natural Sciences,
2-21-1 Osawa, Mitaka, Tokyo 181-8588, Japan}
\affiliation{Department of Astronomical Science,
SOKENDAI (The Graduate University for Advanced Studies),
2-21-1 Osawa, Mitaka, Tokyo 181-8588, Japan}

\begin{abstract}
We present the results of a single-pointing survey of 207 dense cores embedded in Planck Galactic Cold Clumps distributed in five different environments ($\lambda$ Orionis, Orion A, B, Galactic plane, and high latitudes) to identify dense cores on the verge of star formation for the study of the initial conditions of star formation. We observed these cores in eight molecular lines at 76-94 GHz using the Nobeyama 45-m telescope. We find that early-type molecules (e.g., CCS) have low detection rates and that late-type molecules (e.g., {\nth}, {\cch}) and deuterated molecules (e.g., {\ntd}, DNC) have high detection rates, suggesting that most of the cores are chemically evolved. The deuterium fraction (D/H) is found to decrease with increasing distance, indicating that it suffers from \textbf{differential beam dilution between the D/H pair of lines for distant cores ($>$1 kpc).} For $\lambda$ Orionis, Orion A, and B located at similar distances, D/H is not significantly different, suggesting that there is no systematic difference in the observed chemical properties among these three regions. We identify at least eight high D/H cores in the Orion region and two at high latitudes, which are most likely to be close to the onset of star formation. There is no clear evidence of the evolutionary change in turbulence during the starless phase, suggesting that the dissipation of turbulence is not a major mechanism for the beginning of star formation as judged from observations with a beam size of 0.04 pc.
\end{abstract}


\section{Introduction} \label{sec:intro}
In the process of star formation, the initial condition is believed to determine what kinds of stars will be born and how they will form \citep[e.g.,][]{Bergin:2007iy,McKee:2007bd}. However, the details of initial conditions have not been thoroughly characterized.

Stars are formed through gravitational collapse in dense molecular cores \citep[e.g.,][]{Bergin:2007iy,McKee:2007bd}, and many observations have reported that the core mass function is similar in shape to the initial mass function \citep[e.g.,][]{Andre:2010ka, Konyves:2010ff}. However, this is not the case in high-mass regimes, where deviations from the Salpeter IMF have been recently found \citep[e.g.,][]{zhang:2015,Motte:2018ig,Sanhueza:2019cs}. These studies suggest that studying dense cores in different mass ranges may provide valuable information on the initial conditions of star formation. Furthermore, we need to take care of the evolution in the starless phase 
because the internal structure of dense cores is altered physically and chemically during the core evolution \citep{Aikawa:2008dm}. Therefore, the identification of the dense core \textit{just prior to} the beginning of gravitational collapse, which we call `{\pscore}' here, is essential in understanding the genuine initial conditions of star formation \citep[e.g.,][]{Ohashi:2018jr}.
 
In low-mass star-forming regions, prestellar cores are not only compact ($\la$ 0.1 pc), cold ($\la$ 10 K), and dense ($>5\times10^{4}$ cm$^{-3}$) \citep[e.g.,][]{Bergin:2007iy, McKee:2007bd}, but also gravitationally bound, thermally supported, and centrally concentrated \citep[e.g.,][]{WardThompson:1994tg,Caselli:2011hd}. For high-mass star formation, prestellar cores are as massive as at least 30 {\msun}, assuming a star formation efficiency of 30{\%} \citep{wang:2014, Sanhueza:2019cs}.
 
The chemical characteristic of starless cores close to the onset of star formation is high deuterium fractionation. The deuterium fraction (D/H, hereafter) is defined as a column-density ratio of a deuterated molecule to its hydrogenated counterpart, and has commonly been used to study the chemical properties of dense cores in low-mass star-forming regions as well as in high-mass star-forming regions \citep[e.g.,][]{Hirota:2001fq, Crapsi:2005kp, Hirota:2006fh, Bergin:2007iy, Chen:2010hn, Chen:2011bo, Fontani:2011ie, Sakai:2012eh,Sakai:2015eh, Sakai:2018db, Feng:2019bm}.  

The chemical evolution of a dense core in terms of D/H can be described as follows. As a dense core evolves toward gravitational collapse, it develops a steep radial density profile that increases toward the core center and a radial temperature profile that decreases toward the core center \citep[e.g.,][]{Crapsi:2007ge,Aikawa:2008dm}. At the core center ($\la$ 5000 AU) with low temperature ($T_{\rm dust}\lesssim$ 25 K and often closer to 10 K) and high density ($n_{\rm H_2} \geq 10^4$ cm$^{-3}$), CO molecules (a destroyer of deuterated molecules) become frozen on the surface of the dust grain and the enrichment of deuterated molecules is activated with exothermic ion-molecule reactions from HD molecules \citep[e.g.,][]{Millar:1989bb,Phillips:2003vu,Vastel:2004ja, Aikawa:2005kt,Crapsi:2005kp}. When a protostar is born at the core center, it heats up the surroundings \citep[e.g.,][]{Caselli:2002eg}. At the warm ($T_{\rm dust}\gtrsim$ 25 K) core center, CO molecules are desorbed from the grain surface and then destroy deuterated molecules \citep{Roberts:2000un,Lee:2004fi}. Therefore, D/H varies with temperature. If the core is cold enough ($T_{\rm dust}<25$ K), D/H increases in the starless phase and reaches its maximum at the onset of star formation, and then decreases by an order of magnitude after star formation \citep[e.g.,][]{Hirota:2003bk,Crapsi:2005kp, Emprechtinger:2009bp, Friesen:2010hm,Sakai:2012eh,Fontani:2014jy}.

$N$(DNC)/$N$(HNC) and $N$({\ntd})/$N$({\nth}) can be used as D/H tracers. {\ntd} is known to be the least depleted molecule \citep{Pagani:2007fl} and DNC is known to trace cold molecular gas \citep{Gerner:2015ht}. The destruction timescale of the DNC molecule is known to be much longer than that of the {\ntd} molecule \citep[e.g.,][]{Sakai:2012eh,Fontani:2014jy}. For example, after stellar birth, $N$({\ntd})/$N$({\nth}) decreases within 100 years, whereas $N$(DNC)/$N$(HNC) drops at a timescale of 10$^4$ years. \citet{Fontani:2014jy} showed that the deuterium fractionation of {\nth} is more suitable than that of HNC to identify cores on the verge of star formation in high-mass star-forming regions.

For the chemical evolution of the dense core, it is also noticed that the  abundance of gas-phase molecules decreases at different timescales because of adsorption and/or chemical reactions in the central dense region of the core \citep{Aikawa:2001bw,Aikawa:2003ek,Lee:2003hy}. For example, carbon-chain molecules (e.g., CCS, {\hcn}) are abundant in starless cores, whereas {\cch} and nitrogen-bearing molecules (e.g., {\nth}, NH$_3$) are abundant in protostellar cores \citep[e.g.,][]{Suzuki:1992gg, Benson:1998gi, Ohashi:2014hj, Ohashi:2016fo}. That is, CCS molecules are known as  `early-type species' and {\cch}, {\nth}, and NH$_3$ molecules are known as `late-type species'. This means that \textit{chemically evolved} cores have lower abundances of C-chain molecules and higher abundances of {\cch} and N-bearing molecules. $N$(\nth)/$N$(CCS) and $N$({\nth})/$N$({\hcn}) can be valid chemical evolution tracers at a spatial size scale of $\sim$0.1 pc for cold ($T_{\rm dust}\lesssim$ 25 K) cores. These ratios are invalid as evolution tracers for warm ($>$ 25 K) cores, because the {\nth} molecule is destroyed by CO desorption \citep[e.g.,][]{Aikawa:2001bw, Aikawa:2003ek, Lee:2003hy, Tatematsu:2014fx}.

Measurements of the column-density ratios of deuterated/hydrogenated molecules and N-bearing/C-chain molecules provide a useful tool to characterize the evolutionary status of the dense core in which a star will form \citep[e.g.,][]{Hirota:2006fh}. In this paper, {\hcn} is regarded as a C-chain molecule rather than an N-bearing molecule. Taking chemical evolution into account, \citet{Tatematsu:2017hm} introduced the Chemical Evolution Factor (CEF, hereafter) as a comprehensive diagnostic tool for the degree of evolution of dense cores. The CEF is defined so that, when a dense core evolves, the CEF monotonically increases from ${\sim}-$100 for a starless core to $\sim$100 for a protostellar core through zero corresponding to the onset of star formation (See Section \ref{sec:calcef}). The first attempt was made toward low-mass starless cores in nearby cold clouds in the literature \citep{Tatematsu:2017hm}, but the CEF should be established for other environments (e.g., distant high-mass starless cores) by increasing sample numbers.

Efforts toward the identification of dense cores over all sky have extended from optical to infrared wavelengths with instrumental development \citep[e.g., dark cloud in DSS, infrared dark cloud in MSX and Spitzer,][]{Lee:1999fi, Simon:2006ha, Peretto:2009gh, Kim:2010dr,Dobashi:2011ua}. Most recently, the \textit{Planck} all-sky survey has been made and provides a catalog of Planck Galactic Cold Clumps (PGCC, hereafter) \citep{Collaboration:2011js, Collaboration:2016dm}, which includes 13,188 sources of $\sim$5-10 arcmin-sized structures ``clumps'' with temperatures of 10-20 K in various environments. An unbiased selection of PGCCs was mapped with \textit{Herschel} in far-IR band \citep[e.g.,][]{Juvela:2010ja, Juvela:2012fea, Juvela:2018jz}. A series of follow-up observations have been conducted with ground-based radio telescopes such as PMO, CSO, SMT, APEX, NANTEN2, IRAM, Mopra, Effelsberg, SMA, JCMT, and TRAO, and are summarized by \citet{Liu:2015iba,Liu:2018ii}. Specifically, the JCMT large program ``SCUBA-2 Continuum Observations of Pre-protostellar Evolution (SCOPE)'' identified $\sim$20-60 arcsec-sized structures or ``cores'' within PGCCs \citep{Liu:2018ii}. 
These PGCC cores are distributed in low/high-mass star-forming regions from the Galactic plane to high latitudes \citep{Eden:2019ia}. The initial conditions of star formation in these various environments are still poorly understood. For example, the initial conditions of high-mass star formation are poorly known and debatable because of the small numbers of massive prestellar cores in high-mass star-forming regions so far identified \citep[e.g.,][]{wang:2014, Contreras:2018dm, Sanhueza:2017bd, Sanhueza:2019cs} because of their large distances and short evolutionary timescale \citep{McKee:2007bd, Caselli:2011hd}. These PGCC cores can be valuable targets not only to assess evolutionary stages on the basis of the column-density ratios of D/H and N-bearing/C-chain molecules, but also to study the initial conditions of star formation in widely different environments.

Our ultimate goal is to explore the details of the initial conditions for star formation based on the physical and chemical properties of {\pscore}s in various environments. The first step is to identify reliable samples of such dense cores. Thus, we report the first stage of our research with single-pointing observations of 207 cores in PGCCs in the molecular lines of {\ntd} $J = 1-0$, {\nth} $J = 1-0$, DNC $J = 1-0$, {\hnc} $J = 1-0$, CCS $J_N = 7_6-6_5$, CCS $J_N = 8_7-7_6$, {\hcn} $J = 9-8$, and {\cch} $J_{K_a K_c} =  2_{12}-1_{01}$ with the 45-m single-dish telescope of the Nobeyama Radio Observatory\footnote{Nobeyama Radio Observatory is a branch of the National Astronomical Observatory of Japan, National Institutes of Natural Sciences.}. Cores found in this study are in the chemically evolved stage of starless cores ({\lscore}, hereafter) as prestellar core candidates. In the next paper, we will investigate whether these {\lscore}s are actually {\pscore}s or not through mapping observations. 

This study is organized as follows: observations and data analysis procedures are described in Section \ref{sec:obs}, the observational results are presented in Section \ref{sec:res}, Section \ref{sec:dis} presents the identification of {\lscore}s in various environments using the CEF and D/H, and the study is summarized in Section \ref{sec:sum}. 

\begin{figure}[t!]
\figurenum{1}
\epsscale{1.1}
\plotone{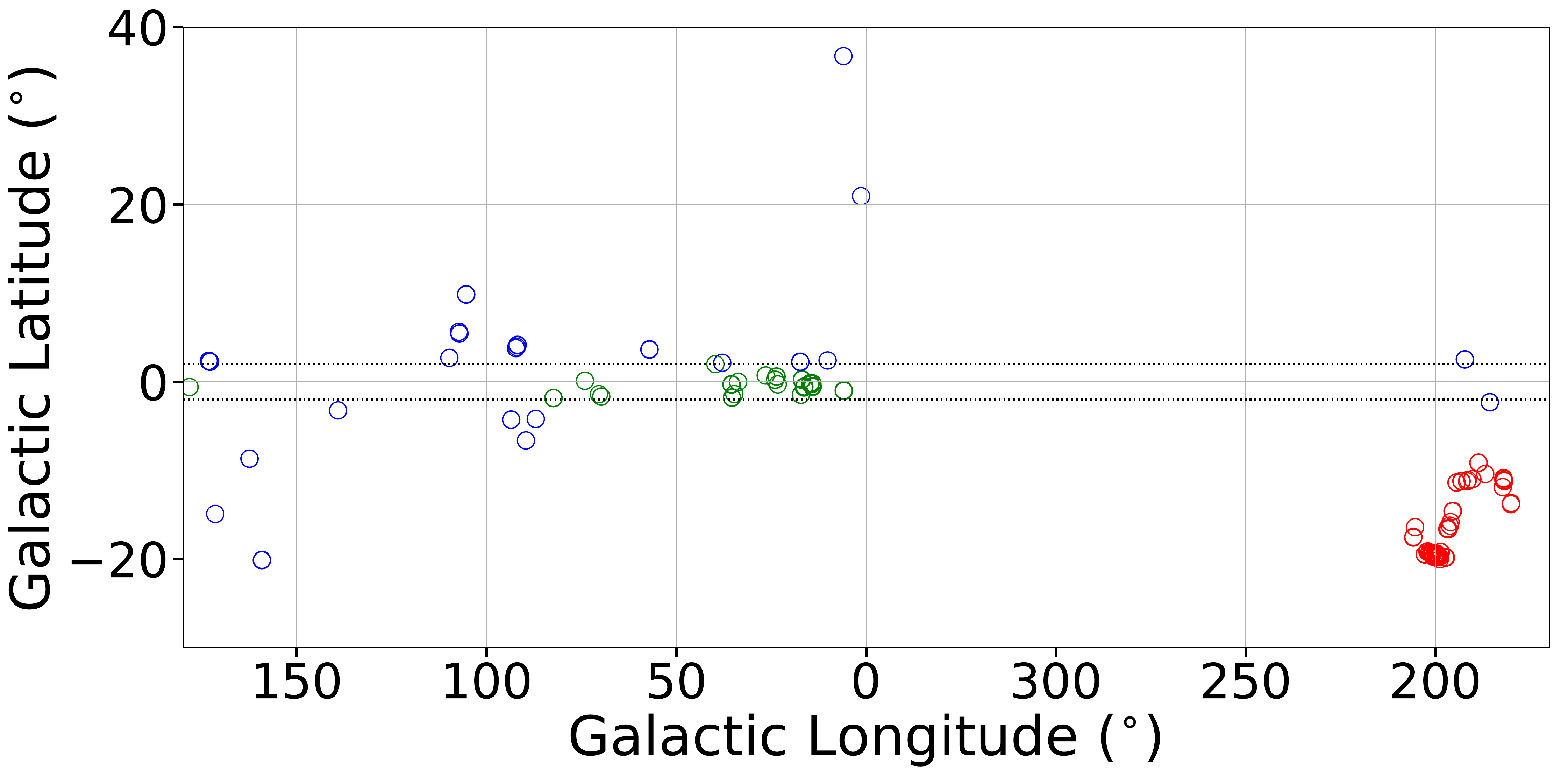}
\caption{Galactic distribution of 207 SCUBA-2 cores. They consist of GMC cores in the Orion region (red), cores in the Galactic plane (green), and cores at high latitudes (blue). The dotted line represents Galactic latitudes of $\pm$2{\arcdeg}. \label{fig:obj_lb}}
\end{figure}

\begin{figure}[t!]
\figurenum{2}
\epsscale{0.8}
\plotone{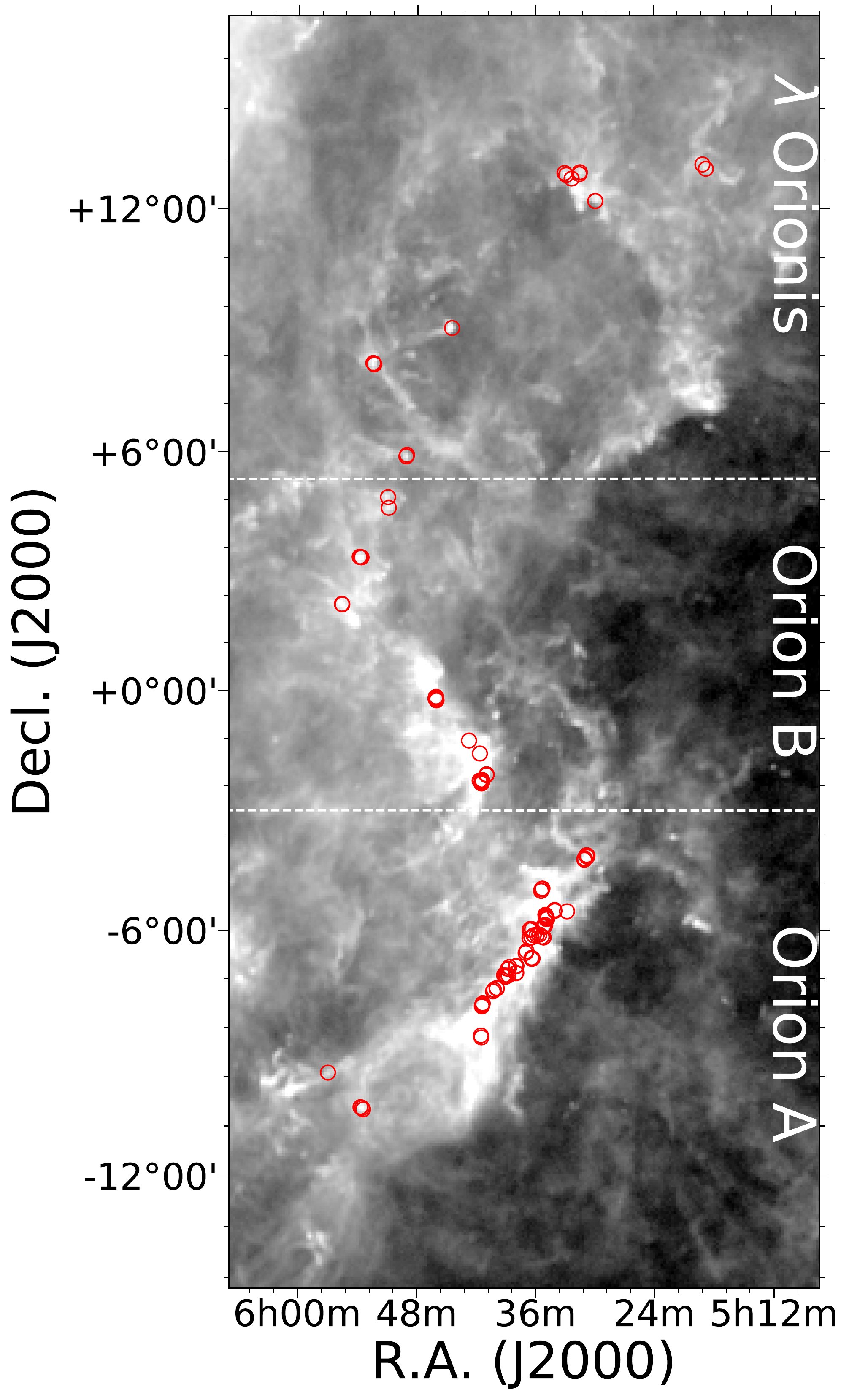}
\caption{Spatial distribution of SCUBA-2 cores in the Orion region. The background image is the \textit{Planck} dust continuum map at 850 {\micron} with a size of 15{\arcdeg}$\times$30{\arcdeg}. The symbol is the same as those used in Figure \ref{fig:obj_lb}. The horizontal lines represent the boundaries for three sub-regions ($\lambda$ Orionis, Orion A, and Orion B).  \label{fig:obj_ori}}
\end{figure}

\section{Observations and data analysis procedure} \label{sec:obs}
\subsection{Sample selection}
The SCOPE survey discovered more than 3,000 PGCC cores across the Galaxy at 850 {\micron} with SCUBA-2 (Submillimetre Common-User Bolometer Array 2) on board JCMT \citep[e.g.,][]{Yi:2018hc,Eden:2019ia}. \citet{Yi:2018hc} identified 119 cores embedded in 96 PGCCs in the Orion region from JCMT SCOPE data and JCMT archive data. \citet{Eden:2019ia} released a compact source catalog using the JCMT SCOPE data, which contains 3,528 cores identified in 1,235 PGCCs from the Galactic plane to high latitudes. As a pilot study, we selected a total of 207 targets that are located at different Galactic positions (local Giant Molecular Clouds (GMCs), high latitude, and Galactic plane). These dense cores are among the densest cores ($N$(H$_2$) $> 1\times10^{22}$ cm$^{-2}$) discovered in the SCOPE survey and consist of 113 cores in the Orion region, which is associated with GMCs, 52 cores in the Galactic plane ($|b|<2\arcdeg$), and 42 cores at high latitudes ($|b|\geq2\arcdeg$). Figure \ref{fig:obj_lb} shows the Galactic distribution of 207 SCUBA-2 cores, and the information on the targets is summarized in Table \ref{tbl:target} with the core name, coordinate, dust temperature, H$_2$ column density, type of environment, and the related PGCC. The dust temperature is taken from the PGCC catalog and the H$_2$ column density is derived from the 850 {\micron} peak intensity of the PGCC core \citep{Yi:2018hc, Eden:2019ia} and the dust temperature. The 207 SCUBA-2 cores have dust temperatures ranging from 9.2 K to 22.4 K with the median value of 13.3 K. They have H$_2$ column densities of (0.17-12)$\times10^{23}$ cm$^{-2}$ with the median value of 8.2$\times10^{22}$ cm$^{-2}$. 

We investigate environments surrounding SCUBA-2 cores using the SIMBAD database. The Orion region, including the 113 SCUBA-2 cores, is a local active star-forming region where both massive stars and low-mass stars are born \cite[e.g.,][]{Yi:2018hc}. This region consists of $\lambda$ Orionis, Orion A, and B sub-regions, which include GMCs and many star-forming regions such as HII regions (e.g., LDN 1641 S3), star clusters (e.g., Collinder 69, Orion Nebula Cluster, NGC 2023, OMC 2), and nebulas (e.g., Orion KL, McNeil Nebula, IC 432). 15, 70, and 28 SCUBA-2 cores are located in the $\lambda$ Orionis, Orion A, and Orion B, respectively, and their distributions are shown in Figure \ref{fig:obj_ori}. \citet{Yi:2018hc} found that the physical properties of the SCUBA-2 cores in the three sub-regions are different. The SCUBA-2 cores correspond to part of the dark clouds (e.g., L1581, L1582, L1594, L1598, L1630, and L1641).

The 52 SCUBA-2 cores in the Galactic plane are located in the vicinity of open clusters (e.g., NGC2264, NGC6530, NGC6611) and supernova remnants (e.g., W48) and are embedded in dark clouds (e.g., L463). The 42 SCUBA-2 cores at high latitudes seem to be associated with famous star-forming regions (e.g., Auriga, California, Cepheus, Cygnus, IC5146, Ophiuchus, Perseus, and Taurus), nearby dark clouds (e.g., Banard 1, L43, L183, L769, L944, L973, L1004, L1181, L1204, L1035, L1521F, and L1525), open clusters (e.g., NGC2264), and star clusters (e.g., NGC7129).

The information on these environments is listed in Table \ref{tbl:target}. We consider $\lambda$ Orionis, Orion A, B, Galactic plane, and high latitudes as five different environments.

\subsection{Single-pointing observations\label{sec:obs:nro}}
We carried out single-pointing observations toward the 850 $\mu$m intensity peak position of 207 SCUBA-2 cores with the 45-m single-dish radio telescope of the Nobeyama Radio Observatory from 2017 February to 2018 May (CG161004, LP177001, P.I.: K. Tatematsu). Either of the two-sideband SIS receivers, T70 or TZ, was used for simultaneous four-line observations in the double-polarization mode \citep{Asayama:2013dl, Nakajima:2013fn}. The T70 receiver was adopted for observations towards all 207 SCUBA-2 cores in the {\cch} $J_{K_a K_c} =  2_{12}-1_{01}$, DNC $J = 1-0$, {\hnc} $J = 1-0$, and {\ntd} $J = 1-0$ molecular emission lines. The TZ receiver was used for observations towards 111 SCUBA-2 cores in the Orion region in the CCS $J_N = 7_6-6_5$ (CCS-L, hereafter), CCS $J_N = 8_7-7_6$ (CCS-H, hereafter), {\hcn} $J = 9-8$, and {\nth} $J = 1-0$ lines. Observations with the TZ receiver were not conducted for two Orion cores and all cores in environments other than the Orion region due to time limitations. The rest frequencies and their references, upper energy levels, and employed receivers of the eight lines are listed in Table \ref{tbl:line}. At 82 GHz, the half power beam widths (HPBW) of the T70 and TZ receivers are 19{\farcs}5$\pm$0{\farcs}3 and 18{\farcs}8$\pm$0{\farcs}1, respectively. Their main-beam efficiencies $\eta_{mb}$ are 54.7$\pm$3.5{\%} and 55.4$\pm$3.5{\%}, respectively. The FX digital spectrometer SAM45 \citep{Kamazaki:2012jm} was employed with a spectral resolution of 15.26 kHz (corresponding to $\sim$0.06 {\kms} at 82 GHz) for 113 SCUBA-2 cores in the Orion region and 30.52 kHz (corresponding to $\sim$0.1 {\kms} at 82 GHz) for the remaining 94 cores. The use of two different spectral resolutions was accidental. The system temperature was typically 200 K. 

\begin{deluxetable}{lCccc}[!tbp]
\tablecaption{Molecular lines \label{tbl:line}}
\tablewidth{0pt}
\tabletypesize{\scriptsize}
\tablenum{2}
\tablehead{
\colhead{Molecular Line} & \colhead{Freq. (GHz)} & \colhead{Ref.} &
\colhead{E$_u$ (K)} & \colhead{Receiver}
}
\startdata
DNC $J=1-0$ & 76.305717 & 1 & 3.7 & T70\\
{\ntd} $J=1-0$ & 77.109610 & 2 & 3.7 & T70\\
{\cch} $J_{K_a K_c}=2_{12}-1_{01}$ & 85.338906 & 3 & 4.1 & T70\\
{\hnc} $J=1-0$ & 87.090859 & 4 & 4.2 & T70\\
CCS J$_N=7_6-6_5$ & 81.505208 & 5 & 15.3 & TZ \\
{\hcn} $J=9-8$ & 81.881461 & 6 & 19.7 & TZ \\
{\nth} $J=1-0$ & 93.173777 & 7 & 4.5 & TZ \\
CCS $J_N=8_7-7_6$ & 93.870107 & 8 & 19.9 & TZ \\
\enddata
\tablecomments{Column 1: Name of the molecular line, Column 2: Rest frequency for the molecular line, Column 3: Reference for the employed frequency: 1. \citet{Okabayashi:1993jc}, 2. \citet{Anderson:1977he}, 3. \citet{Thaddeus:1985hw}, 4. \citet{Frerking:1979hk}, 5. \citet{Cummins:1986fc}, 6. \citet{PICKETT:1998jh}, 7. \citet{Caselli:1995dw}, 8. \citet{Yamamoto:1990kq}, Column 4: Upper energy level of the molecular line, Column 5: Receiver used to observe the molecular line.}
\end{deluxetable}

Single-pointing observation for each core typically took $\sim$30 minutes, including the telescope overhead time in the position-switching mode, to achieve 0.09 K rms sensitivity at a resolution of 0.1 km s$^{-1}$. The telescope pointing uncertainty was established as $\la$5{\arcsec} through five-point measurement toward 43 GHz SiO maser sources close to targets every $\sim$1-1.5 hours. The resulting spectrum is expressed in terms of the antenna temperature corrected for atmosphere extinction {\ta} obtained by the standard chopper wheel calibration. For the reduction of raw data, the baselines are subtracted and then the spectra are co-added on the software package ``NEWSTAR''\footnote{https://www.nro.nao.ac.jp/$\sim$nro45mrt/html/obs/newstar/} of the Nobeyama Radio Observatory. The reduced data are transferred into a data format of the GILDAS CLASS program\footnote{https://www.iram.fr/IRAMFR/GILDAS/} and are resampled into a 0.1 {\kms} velocity resolution. 

\subsection{Gaussian or Hyperfine structure fitting to a spectrum \label{fit}}
Figure \ref{fig:spt} shows the spectra of four or eight molecular emission lines for SCUBA-2 cores. {\nth} and {\ntd} lines show the seven components of hyperfine transitions\footnote{https://spec.jpl.nasa.gov/} \citep[e.g.,][]{Caselli:2002eg}, but the other lines generally exhibit a single velocity component. For a further analysis, we make a Gaussian (GA) or hyperfine structure (HFS) fitting to each spectrum using the GILDAS CLASS program. For {\cch}, CCS, and {\hcn} lines having single Gaussian shapes, we apply the GA fitting to them to measure the peak temperature ($T_{\rm peak}$), systemic velocity ($V_{\rm LSR}$), and linewidth ($\Delta v$) in Full Width at Half Maximum (FWHM). DNC and {\hnc} lines look like single Gaussian shapes, but they are known to have four and six components of hyperfine transitions, respectively \citep{van:2009hz}. Because our main purpose is to derive the column density, for simplicity we ignore the hyperfine splitting for DNC and {\hnc}, and apply simple Gaussian fitting. The derived linewidths of both lines are overestimated for this reason. For {\ntd} and {\nth} lines, where the seven components of hyperfine transitions are clearly visible, we apply the HFS fitting to them to obtain the excitation temperature ($T_{\rm ex}$), $V_{\rm LSR}$, $\Delta v$, \textbf{optical depth ($\tau$), and $T_{\rm ant}\tau$.} To obtain $T_{\rm peak}$ and $\Delta v$ of the brightest component (\textit{JF$_1$F}=123-012) of the {\nth} and {\ntd} lines, we apply the GA fitting to the brightest component of the lines. We measure the rms ($\sigma$) noise level from the baseline subtraction procedure for the spectrum. The fitting results are listed in Tables \ref{tbl:n2h} and \ref{tbl:n2d}. We regard the peak temperature higher than 3$\sigma$ as detection for further analysis. Otherwise, we give 3$\sigma$ as an upper limit and consider it as non-detection.

\subsection{Distance of 207 SCUBA-2 cores \label{sec:dist}}
Accurate distances of cores are often unavailable because of the lack of references. When an accurate distance to the parent cloud is known in the literature, we adopt the value. Otherwise, we employ distance from the parallax-based distance estimator of the Bar and Spiral Structure Legacy Survey \citep{Reid:2016gy} on the basis of the systemic velocity (Section \ref{fit}) and the sky position of the core. With other lines, the {\cch} line is basically used for the systemic velocity because the line has the highest detection rate (Section \ref{sec:sec:det}). Distance and its reference are summarized in Table \ref{tbl:target}.

The median values of the employed distances are found to be 380 pc for cores in the $\lambda$ Orionis, 430$^{+0}_{-40}$ pc for cores in the Orion A, 390$^{+30}_{-0}$ pc for cores in the Orion B, 2.3$^{+7.0}_{-1.3}$ kpc for cores in the Galactic plane, and 0.8$^{+1.8}_{-0.7}$ kpc for cores at high latitudes. It is noted that most of the cores in the Galactic plane are distributed farther away than cores at other environments. Orion cores are located at similar distances, but cores at other environments are widely scattered in distance. 

\subsection{Classification of SCUBA-2 cores into starless and protostellar cores}
Each of the 207 SCUBA-2 cores may be either before or after stellar birth, so they need to be classified into two groups: starless cores and protostellar cores. Reliable categorization can be difficult, depending on the availability of protostellar data and references. Therefore, we simply categorize them by visually investigating the existence of a young stellar object (YSO) within the criterion radius centered at the peak position of each core. For example, if no YSO is known within the criterion radius, the core is regarded as a starless core (candidate). If a YSO is located within the criterion radius, we consider the core as a protostellar core (candidate). For simplicity, we refer to starless cores and their candidates as starless cores, and protostellar cores and their candidates as protostellar cores. The criterion radius is adopted from \citet{Yi:2018hc} and \citet{Eden:2019ia}. The information of the YSOs is taken from the SIMBAD database including the past literature and protostar catalogs based on large programs such as \textit{Spitzer}, \textit{WISE}, \textit{Herschel}, and \textit{GAIA} missions \citep{Megeath:2012cn, Povich:2013ga, Dunham:2015dp, Marton:2016ji,Marton:2019ev}.

From the visual inspection, the samples of our SCUBA-2 cores are classified into 58 starless cores and 149 protostellar cores. The 58 starless cores consist of 5 cores in the $\lambda$ Orionis, 24 in the Orion A, 10 in the Orion B, 13 in the Galactic plane, and 6 at high latitudes. Out of 149 protostellar cores, 10, 46, 18, 39, and 36 cores are located in the same five categories of environments, respectively. The classification and the associated YSOs are summarized in Table \ref{tbl:target}, but may be changed if the YSO information is updated. Considering this uncertainty, among the total 207 SCUBA-2 cores, the upper limit of the fraction of starless cores is found to be $\sim$28{\%}.

\subsection{Estimation for the column density and column-density ratio of molecules \label{sec:calNR}}
\begin{figure}[t!]
\figurenum{4}
\epsscale{1.15}
\plotone{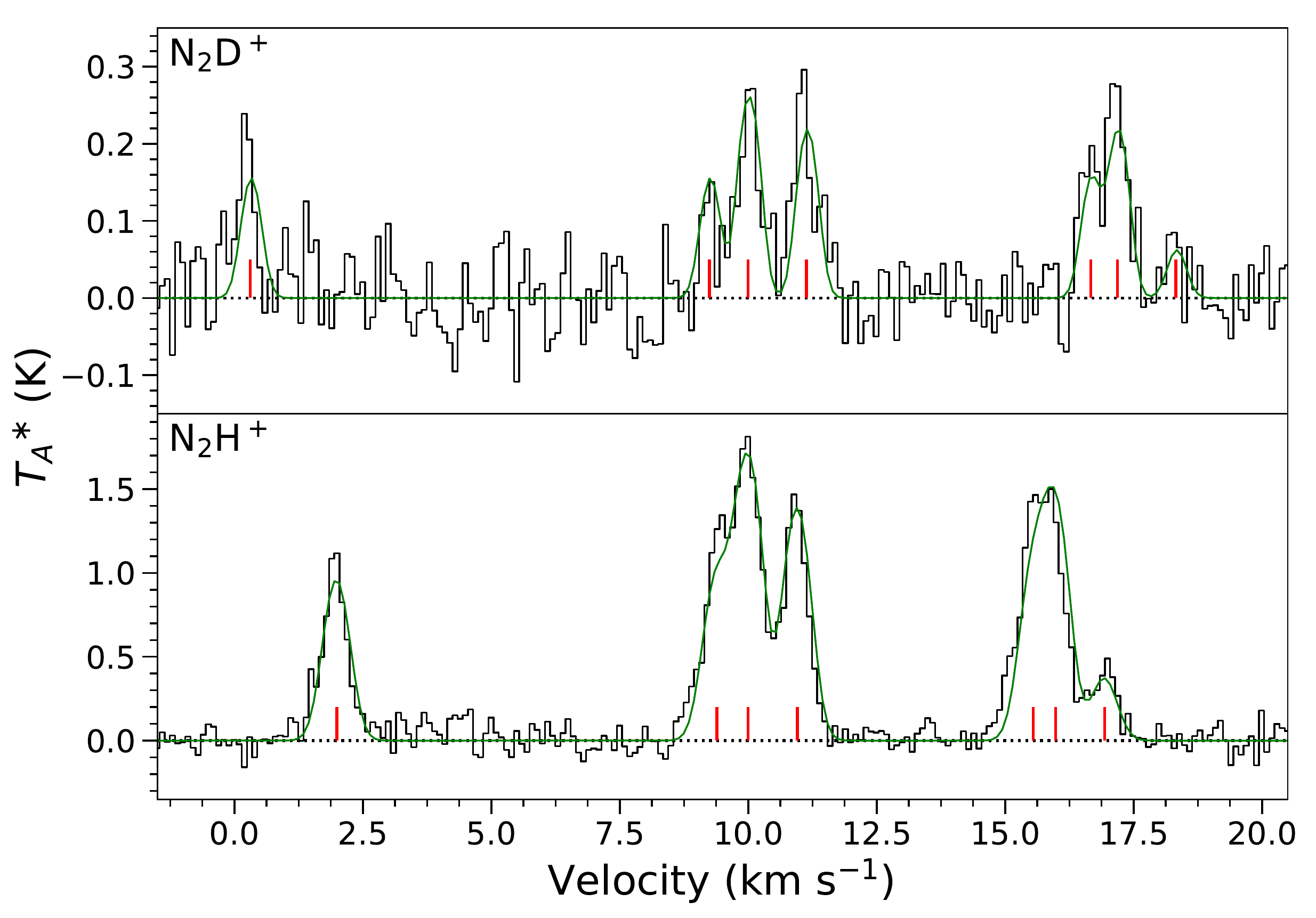}
\caption{\textbf{Examples of excitation anomalies seen in {\ntd} and {\nth} lines. The black and green lines represent the spectrum of G205.46$-$14.56North1 and the result of hyperfine structure fitting, respectively. The red vertical lines represent the velocity offsets of the seven hyperfine components.} \label{fig:hfs_an}}
\end{figure}

We derive the column densities of the six observed molecules, {\ntd}, {\nth}, DNC, {\hnc}, CCS, and {\hcn} (Tables \ref{tbl:n2h} and \ref{tbl:n2d}). Among the two transitions of CCS, we use the lower-frequency transition because it has a higher detection rate (Section \ref{sec:sec:det}). The column densities of the six molecules can be derived with the assumption that the molecular lines are under local thermodynamic equilibrium (LTE) as described in \citet{Suzuki:1992gg}, \citet{Sanhueza:2012iy}, and \citet{Mangum:2015ch}. For {\ntd} and {\nth} lines, the column density is estimated \textbf{through the HFS fitting as follows. (1) When the HFS fitting is successful to some extent, the column density is estimated from $T_{\rm ant}\tau = (T_{\rm ex} - T_{\rm bg})\tau$, as a main factor together with $T_{\rm ex}$ as a weaker contribution. Here, $\tau$ and $T_{\rm bg}$ are the total line optical depth of all the hyperfine components and the temperature of the cosmic microwave background, respectively. (1$-$1) If the HFS fitting is fully successful or the $T_{\rm ant}\tau$ error is $\leq$50{\%} and excitation temperature is in a range of 4 K $\leq T_{\rm ex} \leq$ 25 K, the column density is derived from $T_{\rm ant}\tau$ and $T_{\rm ex}$ from the HFS fitting. (1$-$1$-$1) In a case of the optically thin limit shown as `$\tau=0.1$', $\tau$ is too small to constrain. The relative uncertainty in $T_{\rm ant}\tau$ is small, although that in $\tau$ is large. (1$-$2) If the HFS fitting is partially successful or the error of $T_{\rm ant}\tau$ is $\leq$50{\%} but $T_{\rm ex}$ is out of the above range, $T_{\rm ex}$ is estimated from dust temperature. We assume that gas kinetic temperature ($T_{\rm k}$) is equal to dust temperature, and also assume $T_{\rm ex} = 0.5~T_{\rm k}$ following \citet{Tatematsu:2017hm}, who studied nine Planck Galactic clumps in the NH$_3$ lines. (2) When the HFS fitting is not successful or the $T_{\rm ant}\tau$ error is $>$50{\%}, the column density is derived through single Gaussian fitting to the brightest hyperfine component of the {\ntd} and {\nth} lines. The excitation temperature is, again, estimated from dust temperature. For DNC, {\hnc}, CCS, and {\hcn} lines,} the column density is calculated from the peak temperature \textbf{and linewidth through the Gaussian fitting, and excitation temperature estimated from dust temperature. We estimate the uncertainty of the column density through propagation from the 1$\sigma$ fitting error and the rms noise level of the spectrum. Some cores (e.g., G205.46$-$14.56North1) show excitation anomalies that the ratios of the hyperfine are not fit well with a single excitation temperature (See Figure \ref{fig:hfs_an}). Note that the Rayleigh-Jeans approximation is assumed in the HFS fitting. Such excitation anomalies are reported by \citet{caselli:1995}. We take into account excitation anomalies only as the fitting error in $T_{\rm ex}$. When we employ excitation temperature from dust temperature, we assume that the error in excitation temperature is 50{\%}, following the studies of Tatematsu et al. (\citeyear{tatematsu:2008}, \citeyear{Tatematsu:2017hm}). However, even in this case, when $T_{\rm ex}$ is too low, we can provide only lower limits on the column density. For instance, if $T_{\rm ex}$ is lower than 6.5 K, the line optical depth becomes moderate or large ($\tau > 0.7$) at the lower end of the 50{\%} error range for $T_{\rm ex}$ and we cannot obtain upper limits on the column density reliably. In this case, we show lower limits on the column density obtained at the upper end of the 50{\%} error range for $T_{\rm ex}$.} For undetected lines, we derive upper limits on the column density by using the 3$\sigma$ level at a 0.5 km s$^{-1}$ resolution. The column densities of six molecules are listed in Table \ref{tbl:n_r}.

For the column-density ratios of D/H and N-bearing/C-chain molecules, we calculate $N$({\ntd})/ $N$({\nth}), $N$(DNC)/$N$({\hnc}), $N$({\nth})/$N$(CCS), and $N$({\nth})/$N$({\hcn}) under the assumption that these line emissions are emitted from the same region. If one of the pairing column densities is not estimated, the lower limit or the upper limit is shown. The uncertainty of the column-density ratio is given by propagating from the error of column density in the ratio. Table \ref{tbl:n_r} lists the column-density ratios.  

\subsection{Mach number estimate \label{sec:mach}}
We calculate the Mach number $M=\sigma_{\rm NT}/c_{\rm s}$ to measure the contribution of turbulence in the stability of the dense core. This parameter is used to judge whether a dense core is subsonic ($M\leq$ 1), transonic (1 $<M\leq$ 2), or supersonic ($M>$ 2). The non-thermal velocity dispersion ($\sigma_{\rm NT}$) is derived in the following form \citep[e.g.,][]{Myers:1983it, Fuller:1992bn, Caselli:2002cb}:
\begin{equation}
    \sigma_{\rm NT} = \sqrt{\frac{\Delta v^2}{8~ln~2}-\frac{k_{\rm B} T_{\rm k}}{m}},
\end{equation}
where $k_{\rm B}$ is the Boltzmann constant and $T_{\rm k}$ is the kinetic temperature. $\Delta v$ and \textit{m} are the linewidth (FWHM) and mass of the observed molecule, respectively. The sound speed ($c_{\rm s}$) is estimated by $\sqrt{k_{\rm B} T_{\rm k}/m}$ for the H$_2$ molecule. 

We assume that $T_{\rm k}$ is equal to dust temperature. For $\Delta v$ and \textit{m}, we adopt the {\nth} molecule because it traces the dense gas. The uncertainty of the Mach number is derived by propagating from the error of the fitting, the error of dust temperature, and the rms ($\sigma$) noise level of the spectrum. The Mach number is listed in Table \ref{tbl:n_r}. 

\begin{deluxetable*}{clcccccccc}[t]
\tablecaption{Detection rates of molecular lines in SCUBA-2 cores in different environments \label{tbl:det}}
\tablewidth{0pt}
\tabletypesize{\scriptsize}
\tablenum{6}
\tablehead{
\colhead{Environment} & \colhead{Source} & \colhead{\cch} & \colhead{DNC} & \colhead{\hnc} & \colhead{\ntd} & \colhead{CCS-L} & \colhead{CCS-H} & \colhead{\hcn} & \colhead{\nth} \\
\cmidrule(l{2pt}r{2pt}){3-10}
\colhead{} & \colhead{} & \multicolumn{8}{c}{\%}
}
\startdata
\multirow{3}{*}{All region} & Both (207) & 91 & 78 & 74 & 49 & \nodata & \nodata & \nodata & \nodata \\
& Starless core (58) & 81 & 76 & 60 & 47 & \nodata & \nodata & \nodata & \nodata \\
& Protostellar core (149) & 95 & 79 & 79 & 50 & \nodata & \nodata & \nodata & \nodata \\ 
\hline
\multirow{3}{*}{$\lambda$ Orionis} & Both (15) & 80 & 67 & 60 & 40 & 13 & 13 & 47 & 80 \\
& Starless core (5) & 40 & 40 & 20 & 20 & 0 & 20 & 20 & 60 \\
& Protostellar core (10) & 100 & 80 & 80 & 50 & 20 & 10 & 60 & 90 \\
\hline
\multirow{3}{*}{Orion A} & Both (70) & 89 & 87 & 74 & 57 & 26 & 16 & 76 & 97 \\
& Starless core (24) & 79 & 88 & 63 & 50 & 13 & 13 & 63 & 96 \\
& Protostellar core (46) & 93 & 87 & 80 & 61 & 33 & 17 & 83 & 98 \\
\hline
\multirow{3}{*}{Orion B} & Both (28) & 96 & 96 & 75 & 54 & 18 & 14 & 71 & 89 \\
& Starless core (10) & 90 & 100 & 60 & 60 & 10 & 20 & 70 & 90 \\
& Protostellar core (18) & 100 & 94 & 83 & 50 & 22 & 11 & 72 & 89 \\
\hline 
\multirow{3}{*}{Galactic plane} & Both (52) & 90 & 52 & 67 & 21 & \nodata & \nodata & \nodata & \nodata \\
& Starless core (13) & 85 & 46 & 62 & 23 & \nodata & \nodata & \nodata & \nodata \\ 
& Protostellar core (39) & 92 & 54 & 69 & 21 & \nodata & \nodata & \nodata & \nodata \\
\hline
\multirow{3}{*}{High latitudes} & Both (42) & 98 & 88 & 86 & 69 & \nodata & \nodata & \nodata & \nodata  \\
& Starless core (6) & 100 & 83 & 83 & 83 &  \nodata & \nodata & \nodata & \nodata \\
& Protostellar core (36) & 97 & 89 & 86 & 67 & \nodata & \nodata & \nodata & \nodata
\enddata
\tablecomments{`CCS-L' and `CCS-H' represent the low and high transitions of CCS line, respectively. The numbers in parentheses indicate the total number of cores. The numbers represent the number of dense cores detected in the molecular line in percentages. All detection rates are inferred for sources whose peak temperature is higher than 3$\sigma$. There is no available data of CCS, {\hcn}, and {\nth} lines for cores at other environments except for the Orion region.}
\end{deluxetable*}

\section{Results} \label{sec:res}
\subsection{Detection rate of molecular emission lines} \label{sec:sec:det}
Table \ref{tbl:det} summarizes the detection rates of eight molecular lines in SCUBA-2 cores in different environments. Regarding the detection rates of {\cch}, DNC, {\hnc}, and {\ntd} lines observed towards all SCUBA-2 cores, it is found that the detection rates of all lines are higher than 49{\%}. In particular, the {\cch} (late-type in the cloud chemistry) line shows the highest detection rate (91{\%}) and is detected in almost all cores at high latitudes. The {\hnc} line is also detected at the highest rate in cores at high latitudes. The deuterated molecular lines DNC and {\ntd} are detected at the highest rate in cores in the Orion B and at high latitudes, respectively. All four lines tend to be more detected in protostellar cores than in starless cores

Regarding the detection rates of CCS-L/H, {\hcn}, and {\nth} lines that were only observed towards cores in the $\lambda$ Orionis, Orion A, and B, it is found that, overall speaking, the {\nth} (late-type molecule) and {\hcn} lines show high detection rates ($\geq$47{\%}), whereas CCS lines, which trace the early phase in the cloud chemistry, show low detection rates ($\leq$26{\%}). Among the three environments, all four lines are detected the highest in cores in the Orion A, and the CCS-L, {\hcn}, and {\nth} lines tend to be more detected in protostellar cores than in starless cores, whereas the CCS-H line tends to be more detected in starless cores than in protostellar cores.

In summary, both the low detection rate of early-type molecular lines and the high detection rate of late-type molecular lines and deuterated molecular lines suggest that most of the SCUBA-2 cores are chemically evolved. 

\begin{figure*}[ht!]
\figurenum{5}
\epsscale{1.15}
\plotone{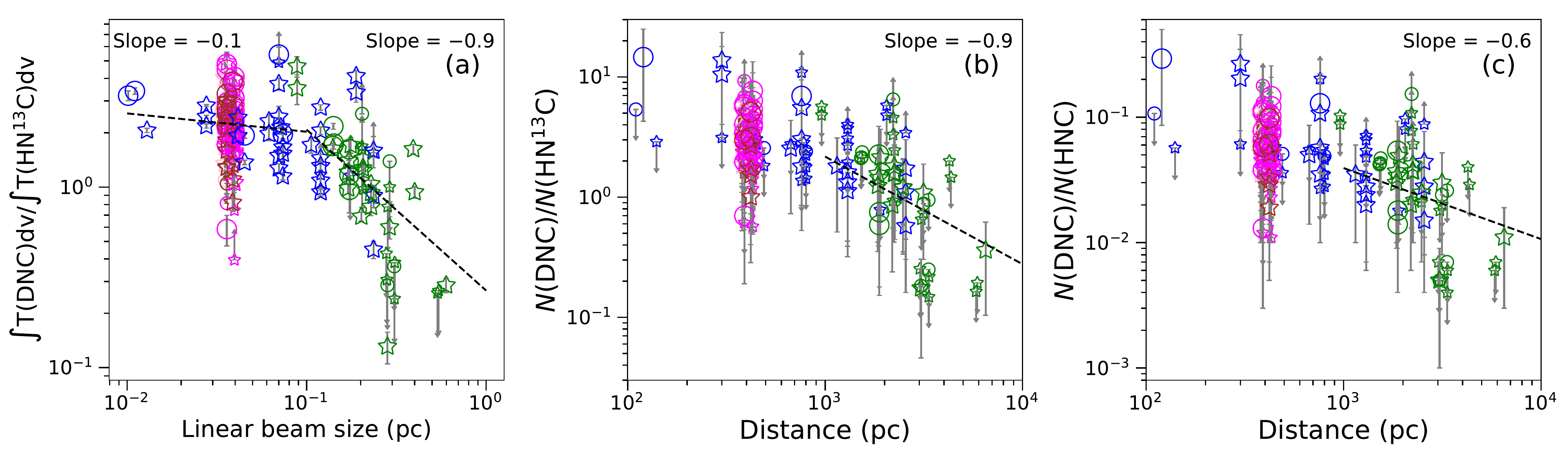}
\caption{\textbf{(a) The integrated-intensity ratio of DNC to {\hnc} is plotted against the linear beam size. (b) The column-density ratio of DNC/{\hnc} is plotted against distance. (c) The same as (b) but for DNC/HNC. The color symbols represent SCUBA-2 cores in the $\lambda$ Orionis (pink), Orion A (magenta), Orion B (brown), Galactic plane (green), and at high latitudes (blue). The circle and star symbols represent the starless and protostellar core, respectively. The large symbol with the error bar represents the dense core where both values in the ratio are successfully estimated. The small symbol with the arrow represents the dense core with upper or lower limits. The dashed line with the slope value represents the least-squares fitting to the cores detected in both molecular lines. In panel (a), the fittings are made for linear beam sizes of $\leq$0.1 pc and of $>$0.1 pc separately. In panels (b) and (c), the fittings are made for distances of $>$1 kpc. } \label{fig:r_d}}
\end{figure*}

\begin{figure*}[ht!]
\figurenum{6}
\epsscale{1.1}
\plotone{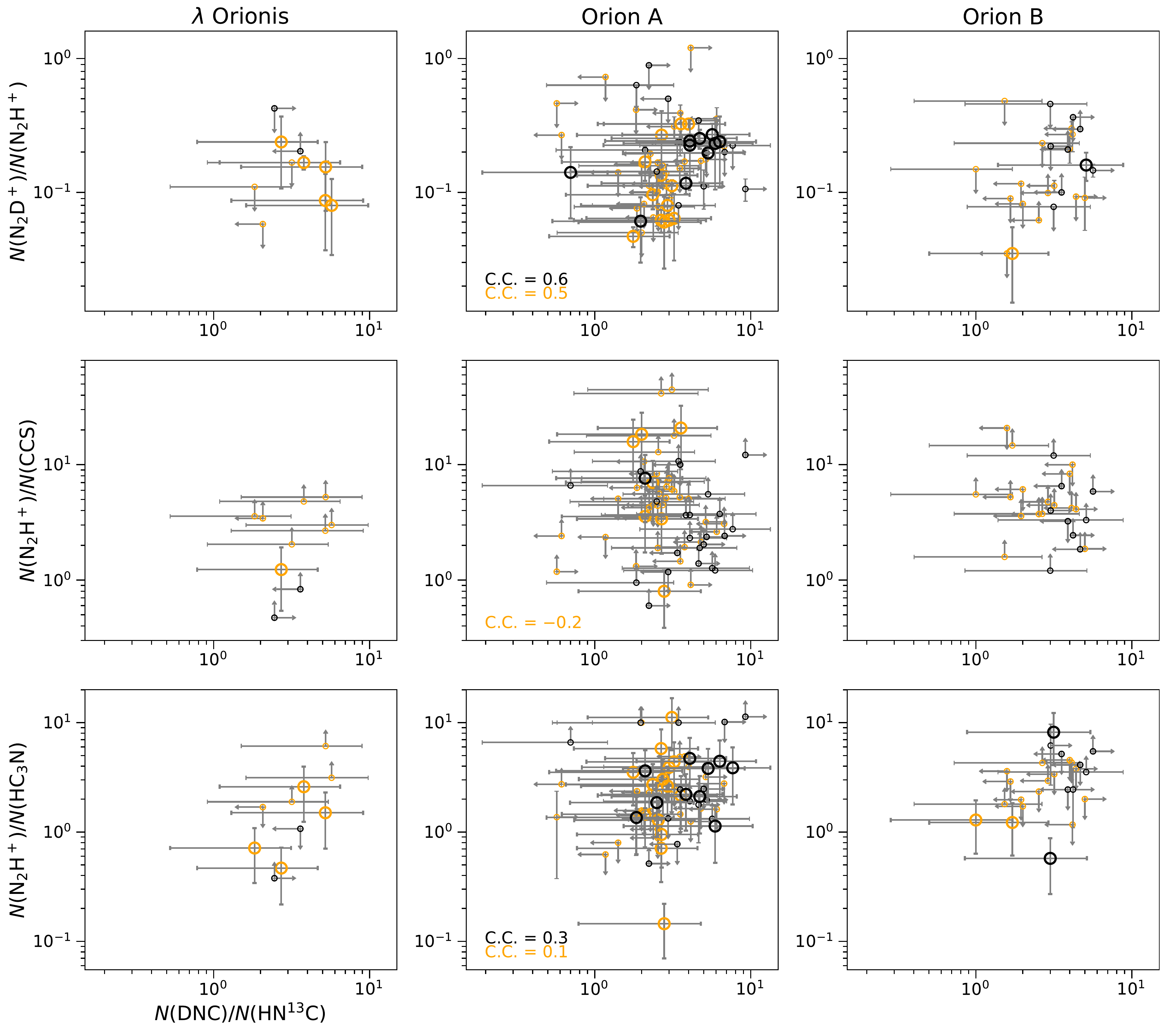}
\caption{The column-density ratios of {\ntd}/{\nth}, {\nth}/CCS, and {\nth}/{\hcn} are plotted against that of DNC/{\hnc} for SCUBA-2 cores in the $\lambda$ Orionis, Orion A, and B. The black and orange represent starless and protostellar cores, respectively. The large symbol with the error bar represents the dense core where both column densities in the ratio are successfully estimated. The small symbol with the arrow represents the dense core with either upper or lower limits. The C.C. represents the correlation coefficient; we show the C.C. only when both the pairing column densities are successfully estimated for seven or more dense cores. The C.C. does not take into account outliers in the sample. \label{fig:corrplot_r}}
\end{figure*}

\begin{figure*}[h!]
\figurenum{7}
\epsscale{1.15}
\plotone{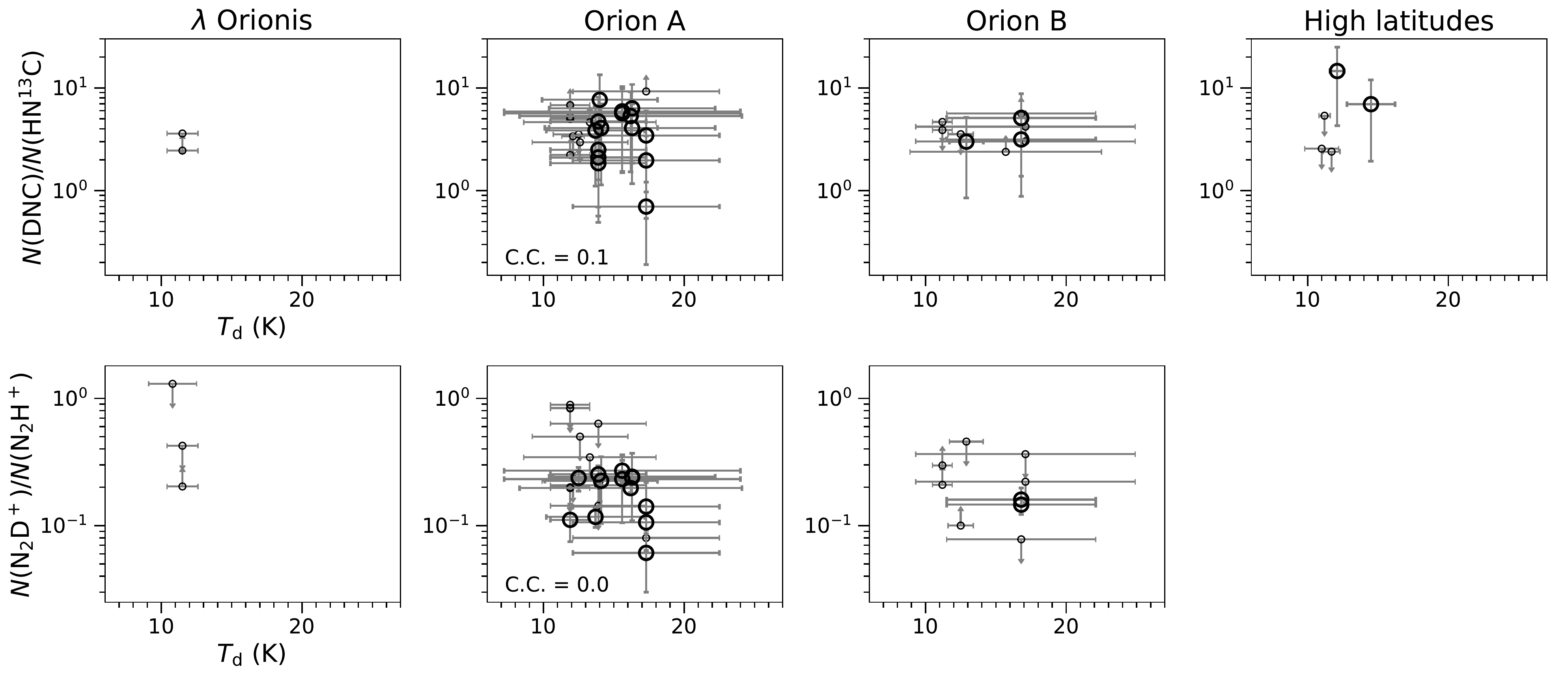}
\plotone{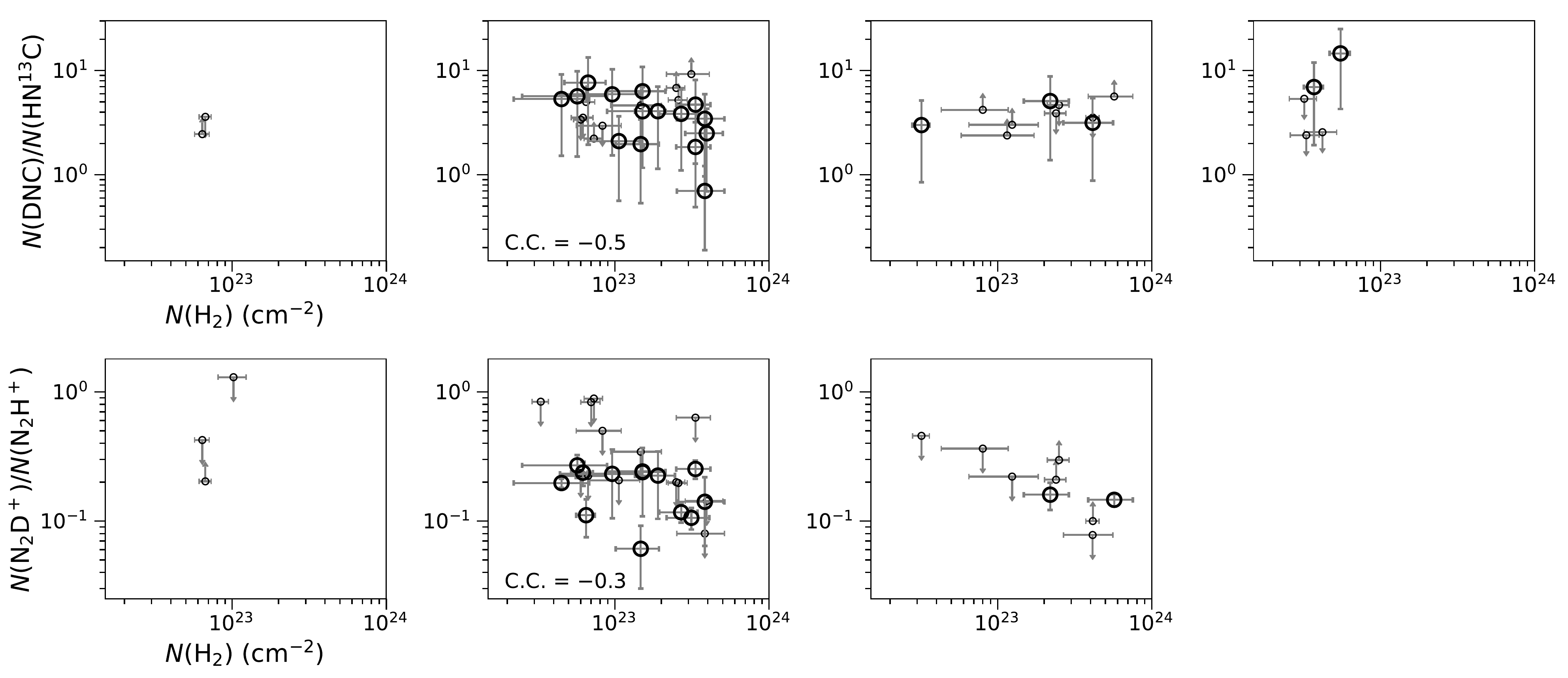}
\caption{The column-density ratios of DNC/{\hnc} and {\ntd}/{\nth} are plotted against the dust temperature and H$_2$ column density for local ($<$1 kpc) starless cores in different environments. The meanings of the symbols and C.C. are the same as those used in Figure \ref{fig:corrplot_r}.  \label{fig:r_td_cdh2}}
\end{figure*}

\begin{figure*}
\figurenum{8}
\epsscale{1.15}
\plotone{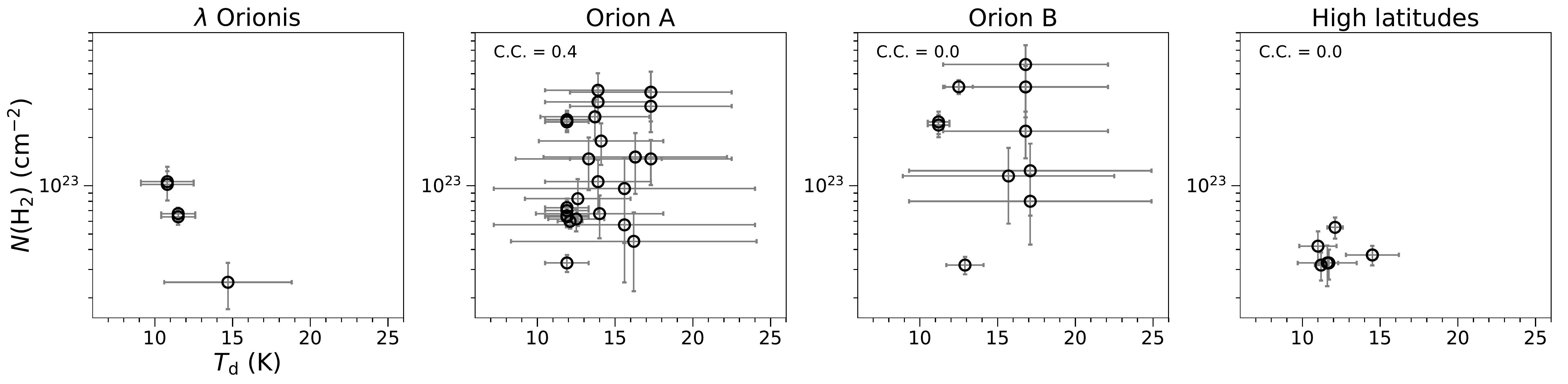}
\caption{The H$_2$ column density is plotted against dust temperature for local ($<$1 kpc) starless cores in different environments. The meanings of the symbols and C.C. are the same as those used in Figure \ref{fig:corrplot_r}.  \label{fig:cdh2_td}}
\end{figure*}

\subsection{Beam dilution effect on distant SCUBA-2 cores \label{sec:r_d}}
Among the four column-density ratios estimated in Section \ref{sec:calNR}, the column-density ratio of DNC/{\hnc} is available for cores in all environments, but the three other column-density ratios, including {\nth} molecules, are available only for cores in the $\lambda$ Orionis, Orion A, and B. \textbf{To estimate the column-density ratio, we assume that the beam filling factor is the same for both the emission lines. However, because the SCUBA-2 cores are located at very different distances ranging from $\sim$0.1 pc to $\sim$10 kpc, our telescope observes them with very different linear beam sizes. Figure \ref{fig:r_d} (a) shows the integrated-intensity ratio of DNC/{\hnc} against the linear beam size in pc. The ratio does not increase or decrease significantly for a beam size of $<$0.1 pc but decreases with increasing beam size for $>$0.1 pc with a power-law index of $-1$. Because dense cores have a typical size of 0.1 pc, distant cores observed with larger beam size can be affected by different beam dilution between the two lines; DNC is beam-diluted while beam dilution for {\hnc} is much weaker. In other words, larger linear beams for distant cores tend to involve the outer, less-dense region having low deuterium fraction. A linear beam size of 19 arcsec corresponds to 0.1 pc at a distance of 1.1 kpc. This differential beam dilution can lead to misinterpretations in the chemical properties of SCUBA-2 cores at different distances. We investigate whether it is the case for the column-density ratio of DNC/{\hnc} and DNC/HNC. In Figure \ref{fig:r_d} (b) and (c), we plot the column-density ratio against distance. The column-density ratio of DNC/HNC is derived by using the abundance-ratio formula of $^{12}$C/$^{13}$C as a function of Galactocentric distance obtained by \citet{savage:2002}. Panels (b) and (c) of Figure \ref{fig:r_d} show similar trends that both the column-density ratios decrease with increasing distance for distant cores ($>$1 kpc). It seems that the column-density ratio is seriously affected by differential beam dilution between the two lines for distant cores (distance $>$ 1 kpc). Therefore, with our beam size, the column-density ratio seems reliable if distance is $<$1 kpc.} 

\textbf{The percentage of protostellar cores (65{\%}) in the Orion region (distance $\sim$ 400 pc) is smaller than that of cores with distances of 2$-$10 kpc in the Galactic plane (81{\%}). It is possible that the protostellar-core percentage increases with increasing distance. It is likely that the YSO sensitivity is shallower for distant cores. Furthermore, the beam dilution may merge weaker starless cores into brighter protostellar cores. In the Orion region, starless cores are half as intense as protostellar cores in the SCUBA-2 flux density. Figure 10 of \cite{Eden:2019ia} shows that the percentage of the SCUBA-2 core detection decreases with increasing distance, which may suggest that the SCUBA-2 sensitivity becomes insufficient for weaker cores at larger distances.}

\textbf{To avoid distance-related issues, we start with the properties of cores in the three sub-regions of the Orion region with similar distances.}

\subsection{Column-density ratios of {\ntd}/{\nth}, DNC/{\hnc}, {\nth}/CCS, and {\nth}/{\hcn} \label{sec:r} for SCUBA-2 cores in $\lambda$ Orionis, Orion A, and B}
Figure \ref{fig:corrplot_r} plots the column-density ratios of {\ntd}/{\nth}, {\nth}/CCS, and {\nth}/{\hcn} against that of DNC/ {\hnc} for starless/protostellar cores in the $\lambda$ Orionis, Orion A, and B. Table \ref{tbl:r_stat} summarizes the statistics of the column-density ratios. We consider only cores where both column densities are successfully estimated. 

For the four column-density ratios, there is no systematic difference between starless and protostellar cores in the $\lambda$ Orionis, Orion A and B. This may indicate that cores in these three regions have similar chemical properties as a whole.

We examine the correlations between the four column-density ratios using the correlation coefficient (C.C) derived by the Pearson product-moment correlation coefficient. This method estimates the strength of the relationship between the relative movements of two variables and provides a value between $-$1.0 and $+$1.0. The values of $+1.0$ and $-1.0$ indicate a strong positive and negative correlation, respectively, and the value of 0 represents no correlation. The C.C. considers only cores of 7 or more in the number of samples where both column densities in the ratio are successfully calculated, but does not take into account outliers in the sample. For $N$({\ntd})/$N$({\nth}) and $N$(DNC)/$N$({\hnc}), cores in the Orion A show a positive correlation. This result suggests that $N$(\ntd)/$N$(\nth) and $N$(DNC)/$N$(\hnc) seem effective for SCUBA-2 cores as chemical evolution tracers.

YSOs inside star-forming cores will have shocks and radiation and make the chemistry of protostellar cores more complex. In this paper, we defer our discussion on protostellar cores to our future papers, and concentrate on the characteristics of starless cores.

\begin{figure*}[th!]
\figurenum{9}
\epsscale{1.15}
\plotone{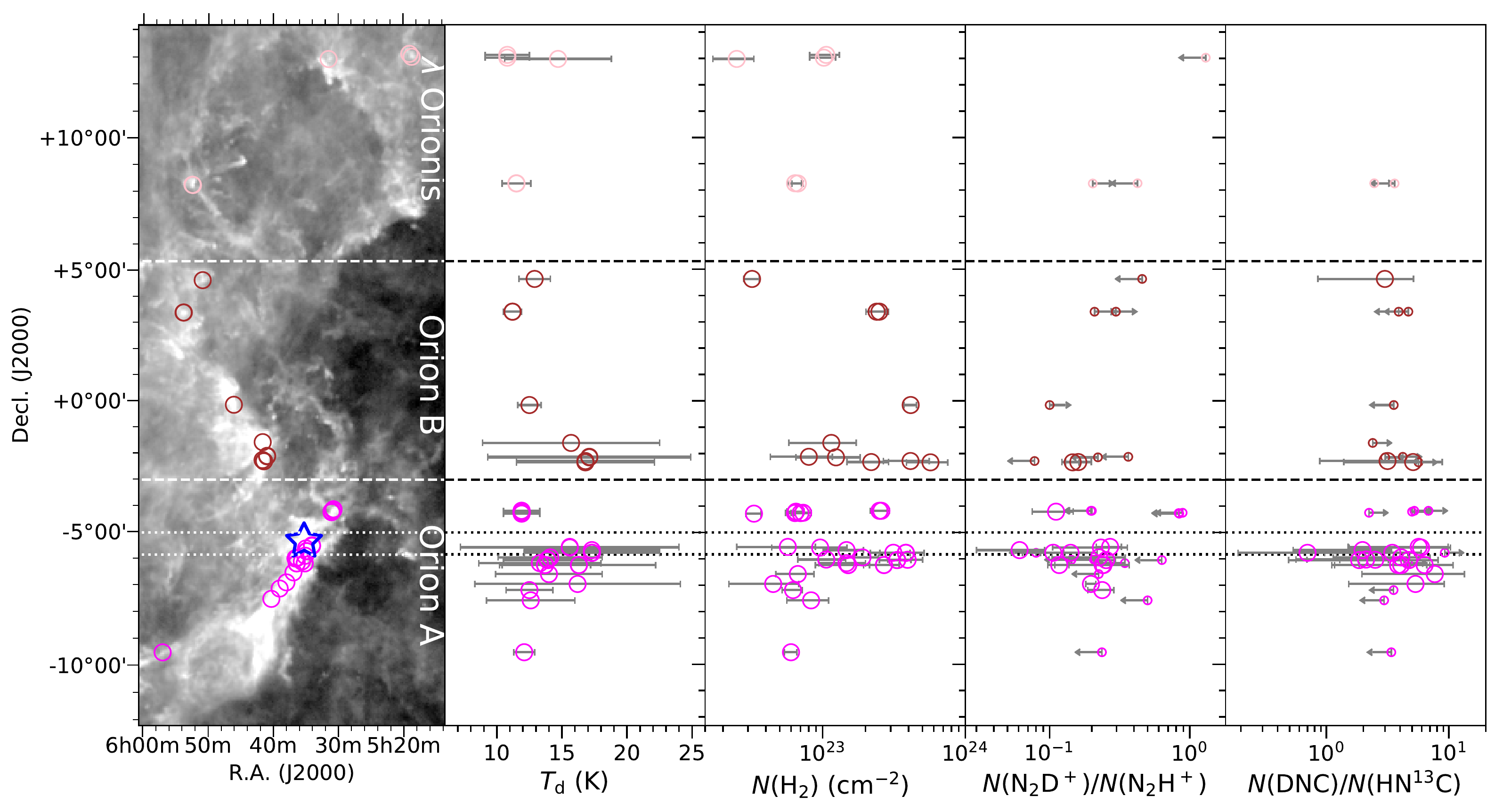}
\caption{Spatial distribution, dust temperature, H$_2$ column density, $N$(\ntd)/$N(${\nth}), and $N$(DNC)/$N(${\hnc}) against declination for starless cores in the Orion region. The large symbol with errorbar and the small symbol with an arrow are the same as those used in Figure \ref{fig:corrplot_r}. The background image in the left panel is the \textit{Planck} 850 {\micron} dust continuum map. The blue star symbol in the left panel represents the position of the Orion Nebula (R.A., Decl. = 05$^h$ 35$^m$ 17.3$^s$, $-$05$\arcdeg$ 23$\arcmin$ 28$\arcsec$). The dashed horizontal lines represent the boundaries of $\lambda$ Orionis, Orion B, and Orion A. The dotted horizontal lines represent the boundaries of the potential impact zone of the Orion Nebula on the declination.  \label{fig:r_ori}}
\end{figure*}

\subsection{Column-density ratio against H$_2$ column density and dust temperature  \label{sec:r_n_dt}}
The H$_2$ column density and dust temperature in the central region of the core can be used as the indicator of core evolution because the density increases and the temperature decreases throughout the starless core evolution \citep{Shirley:2005hy,Aikawa:2008dm}. We investigate whether D/H has any correlation with the dust temperature and H$_2$ column density for local ($<$1 kpc) starless cores in different environments using the correlation coefficient. Figure \ref{fig:r_td_cdh2} shows the column-density ratio of DNC/{\hnc} and {\ntd}/{\nth} against the dust temperature and H$_2$ column density. Cores in the Galactic plane are excluded because all of the cores are more distant than 1 kpc.

For D/H against the dust temperature and against the H$_2$ column density, there seems to be no apparent correlation. This is likely due to the fact that the spatial resolution of the \textit{Planck} telescope (5{\arcmin} corresponds to $\sim$0.5 pc at a distance of $\sim$400 pc) is too low to trace the variation of temperature or density. Figure \ref{fig:cdh2_td} plots the H$_2$ column density as a function of dust temperature for local starless cores in different environments, and shows no clear correlation between the two quantities. \textbf{No clear trend is seen probably because the resolution is too low.}

\subsection{Variation of D/H, dust temperature, H$_2$ column density against declination for starless cores in the $\lambda$ Orionis, Orion A, and B \label{sub:r_sl_ori}}
In the Orion region, shell/ring-shaped structure $\lambda$ Orionis, filamentary clouds Orion A and B are extensively distributed along the declination, as shown in Figure \ref{fig:obj_ori}. For starless cores in the $\lambda$ Orionis, Orion A, and B, we examine whether there is any variation in dust temperature, H$_2$ column density, and D/H along the declination. Figure \ref{fig:r_ori} shows the spatial distribution, dust temperature, H$_2$ column density, $N$({\ntd})/$N$(\nth), and $N$(DNC)/$N$(\hnc) with respect to declination. Along the declination, the dust temperature and H$_2$ column density appear to increase toward a declination of $\sim-4${\arcdeg} to $-3{\arcdeg}$. We exclude cores with a declination from $-5\arcdeg~50\arcmin$ to $-5\arcdeg$ strongly externally heated by the Orion Nebula. \textbf{Although we see slight increase in the dust temperature and H$_2$ column density near the Orion Nebula, we do not see any systematic trend in the D/H fraction.}

\subsection{Comparison of deuterium fraction between starless cores and other cores \label{sec:r_sl_cloud}}
We compare the deuterium fraction between starless cores and other cores reported in previous studies. The average $N$({\ntd})/$N$({\nth}) of starless cores in the Orion region (0.2$\pm$0.1) is found to be lower than that of PGCCs (0.5) reported in \citet{Tatematsu:2017hm}, which is similar to that of low-mass starless cores \citep[0.1,][]{Crapsi:2005kp}, and higher than that of massive protostellar IRAS cores \citep[0.01,][]{Fontani:2006ga}. \textbf{If we assume $^{12}$C/$^{13}$C $=52$ \citep{savage:2002}, the average $N$(DNC)/$N$(HNC) of starless cores in the Orion region (0.08$\pm$0.03)} seems to be similar to those of 13 PGCCs (0.08) found in \citet{Tatematsu:2017hm} and protostellar cores in the Perseus molecular cloud (0.06) in \citet{Imai:2018kz}, but higher than that of low-mass starless cores (0.02) in \citet{Hirota:2006fh}. \textbf{Approximately 30{\%} of the} starless cores in the Orion region are found to have a $N$(DNC)/$N$(HNC) higher than the value of 0.05 for L1544 \citep{Hirota:2003bk, Hirota:2006fh}, suggesting that they could be starless cores more evolved toward the beginning of star formation. 

\section{discussion \label{sec:dis}}

\begin{figure*}[ht!]
\epsscale{1.15}
\figurenum{10}
\plotone{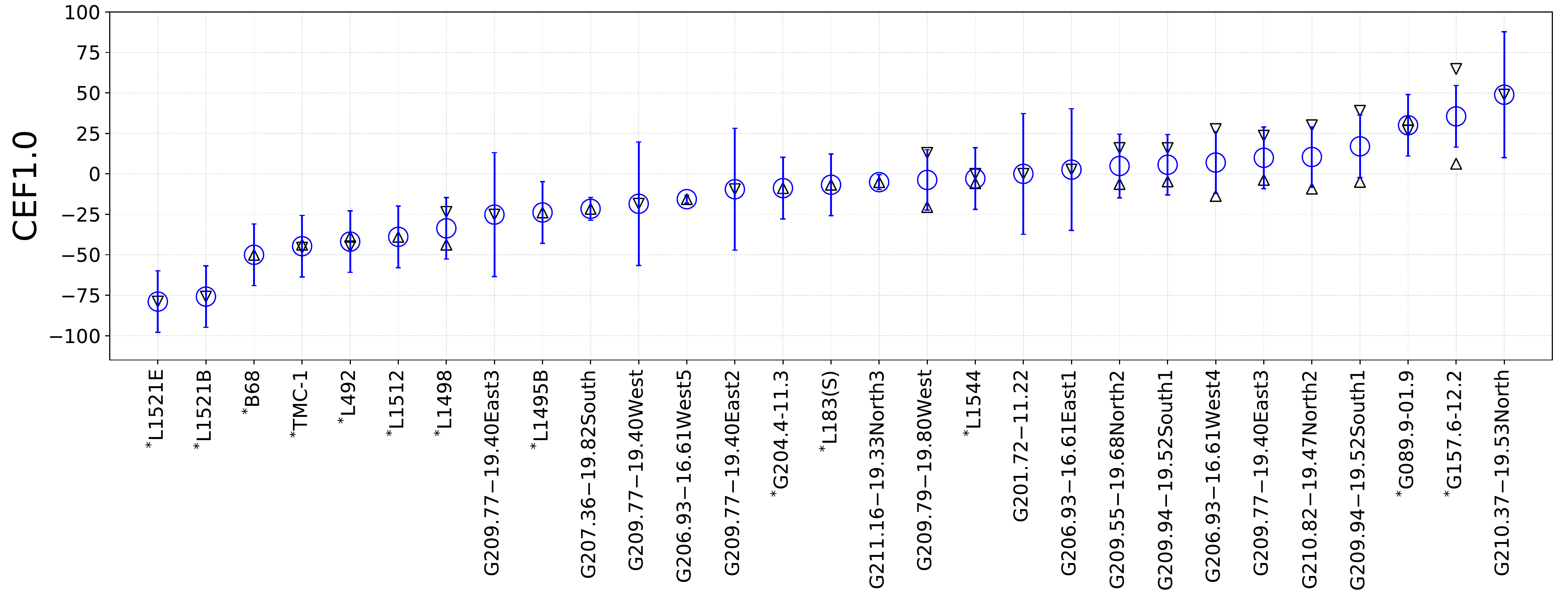}
\plotone{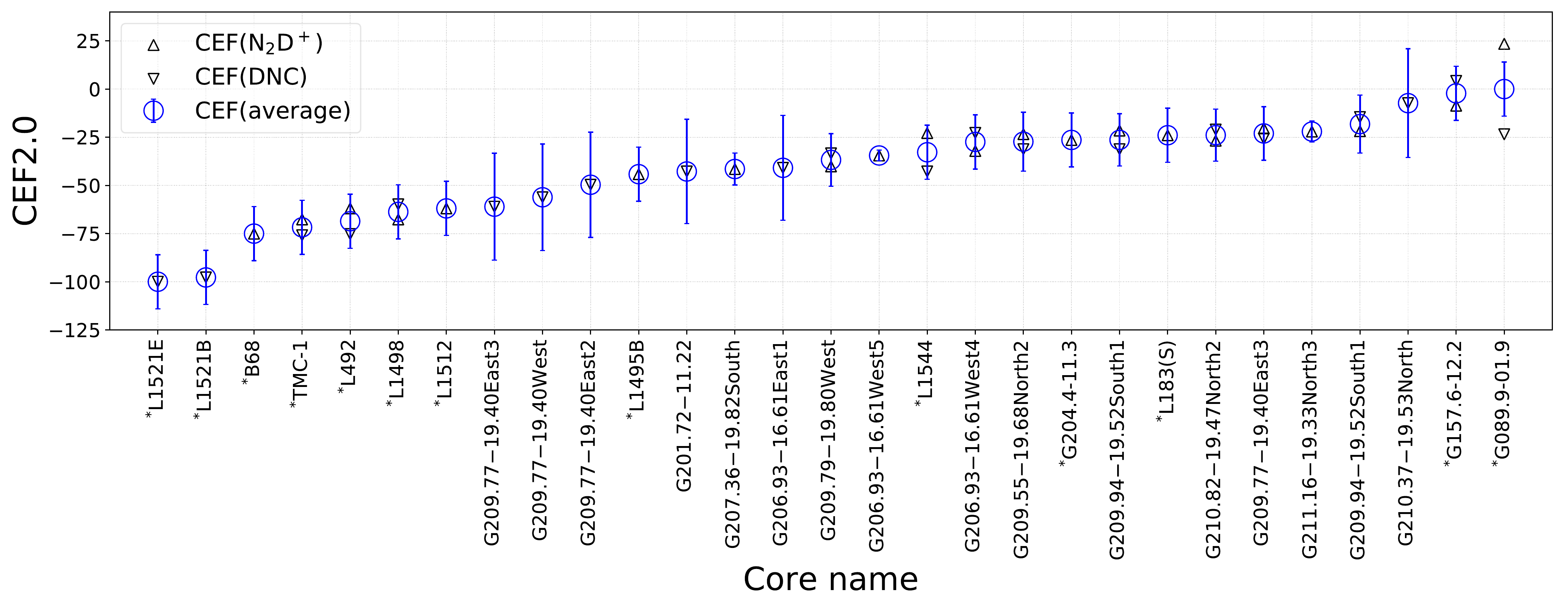}
\caption{Chemical Evolutionary Factor (CEF) for starless cores in the Orion region \textbf{and for local ($<$1 kpc) starless cores studied in \citet{Tatematsu:2017hm} in ascending numeric order} of the average CEF. The top and bottom panels show \textbf{CEF1.0} derived by the equations in \citet{Tatematsu:2017hm} and \textbf{CEF2.0} by our study, respectively. \textbf{The source name with the asterisk represents the cores taken from \citet{Tatematsu:2017hm}.} \label{fig:cef_ori_sl}}
\end{figure*}

\subsection{Identification of the dense cores on the verge of star formation in the Orion region using CEFs \label{sec:calcef}}
The CEF is expressed in the form of ${\rm CEF = log([}N{\rm (A)/}$ $N{\rm (B)]/[}N_0{\rm (A)/}N_0{\rm (B)])\times}d$ for molecules A and B. $N$(A)/$N$(B) is the ratio of column densities inferred from observations, $N_0($A$)/N_0($B$)$ is the ratio of column densities at the onset of star formation. The factor $d$ is defined so that all cores range from ${\sim}-$100 for starless cores to ${\sim}+$100 for star-forming cores through 0 for cores on the verge of star formation. In this paper, we derive the CEF for only starless cores to identify {\lscore}s using $N$(\ntd)/$N$(\nth) and $N$(DNC)/$N$(\hnc). For starless cores, $N_0($A$)/N_0($B$)$ approximately corresponds to the maximum of the column-density ratio of D/H molecules and $d$ is determined for all starless cores so that the CEF ranges from ${\sim}-$100 to $\sim$0. 
Using two column-density ratios listed in Table \ref{tbl:n_r}, we determine $N_0$ and $d$. We consider only starless cores in the Orion region because they should be located at similar distances. Out of them, we consider only cores in which both column densities in the ratio are successfully estimated. We exclude cores externally affected by large-scale star-forming activities (e.g., Orion Nebula, See Figure \ref{fig:r_ori}). \textbf{For the 16 starless cores in the Orion region for which column-density ratios are successfully obtained, we first derive the CEF values using the definition of \citet{Tatematsu:2017hm}, which we call CEF1.0. The top panel of Figure \ref{fig:cef_ori_sl} shows the CEF1.0 values of the Orion cores. We also plot local ($<$1 kpc) cores studied in \citet{Tatematsu:2017hm}. CEF1.0 ranges from ${\sim}-$75 to $\sim$50, which is different from the original definition of the CEF. For example, L1544, known as a gravitational collapsing prestellar core \citep{tafalla:1998}, has a CEF1.0 close to zero, which is reasonable. However, half of the starless Orion cores have positive CEF1.0, suggesting that CEF1.0 of \citet{Tatematsu:2017hm} needs to be updated.}

\textbf{Using both samples of our starless Orion cores and the local cores from \citet{Tatematsu:2017hm}, we determine new CEF2.0 equations in the following forms:}
\begin{align}
&{\rm CEF2.0 (N_2D^+)} = {\rm log}(\frac{N({\rm N_2D^+})/N({\rm N_2H^+})}{0.56})\times59, \\
&{\rm CEF2.0 (DNC)} = {\rm log}(\frac{N({\rm DNC})/N({\rm HN^{13}C})}{9.3})\times87.
\end{align}
\textbf{We search for $N_0($A$)/N_0($B$)$ and \textbf{$d$} for CEF2.0({\ntd}) and CEF2.0(DNC) that minimizes the root mean squares of CEF2.0({\ntd})$-$CEF2.0(DNC) while satisfying a condition that the minimum CEF2.0(average) is $-$100, and the maximum CEF2.0(average) is 0 for the starless cores. The reason for finding  $N_0($A$)/N_0($B$)$ and {$d$} in this manner is to illustrate chemical evolution as simple as possible because different pairs of molecules may show different characteristics. The uncertainty of the CEF is estimated through propagation from the error of the column-density ratio. The derived CEF2.0 is shown in the bottom panel of Figure \ref{fig:cef_ori_sl}, and values for the starless Orion cores are listed in Table \ref{tbl:cef}.}

The \textbf{CEF1.0} of \citet{Tatematsu:2017hm} was constructed on the basis of 15 nearby low-mass starless cores with kinetic temperatures of 10$-$20 K observed at spatial resolutions of 0.015$-$0.05 pc. In the present study, the \textbf{CEF2.0 is constructed by adding the data of the 16 starless GMC cores in the Orion region to the existing samples having similar ranges of temperature and spatial resolution. Part of the origin of differences between the two CEF versions is probably due to the different range of evolutionary stages. For instance, \citet{Tatematsu:2017hm} used cores of a broad range of different environments while the present study investigates one region where environmental differences may be smaller. Indeed, the detection rate of CCS is low in the Orion cores (Section \ref{sec:sec:det}), suggesting that they are chemically evolved. Furthermore, we used SCUBA-2 cores for observations, which are possibly biased to evolved cores having steeper radial intensity distribution. It is likely that these facts make the CEF2.0 values of starless Orion cores closer to zero. Probably our new CEF2.0 can be used more reliably if the telescope beam size is $\lesssim$0.1 pc.} 

\begin{deluxetable}{lCCC}
\tablecaption{Chemical Evolutionary Factor \textbf{(CEF2.0)} for starless cores in the Orion region. \label{tbl:cef}}
\tablewidth{0pt}
\tabletypesize{\scriptsize}
\tablenum{8}
\tablehead{
\colhead{SCUBA-2 core} & \colhead{CEF2.0($\rm N_2D^+$)} & \colhead{CEF2.0($\rm DNC$)} & 
\colhead{CEF2.0(average)} 
}
\startdata
G209.77$-$19.40East3 & \nodata & -61\pm28 & -61\pm28 \\
G209.77$-$19.40West & \nodata & -56\pm28 & -56\pm28 \\
G209.77$-$19.40East2 & \nodata & -50\pm27 & -50\pm27 \\
G201.72$-$11.22 & \nodata & -43\pm27 & -43\pm27 \\
G207.36$-$19.82South & -41\pm8 & \nodata & -41\pm8 \\
G206.93$-$16.61East1 & \nodata & -41\pm27 & -41\pm27 \\
G209.79$-$19.80West & -40\pm4 & -33\pm27 & -37\pm14 \\
G206.93$-$16.61West5 & -34\pm3 & \nodata & -34\pm3 \\
G206.93$-$16.61West4 & -32\pm6 & -23\pm28 & -27\pm14 \\
G209.55$-$19.68North2 & -23\pm14 & -31\pm27 & -27\pm15 \\
G209.94$-$19.52South1 & -21\pm3 & -31\pm27 & -26\pm14 \\
G210.82$-$19.47North2 & -27\pm2 & -21\pm27 & -24\pm14 \\
G209.77$-$19.40East3 & -20\pm4 & -26\pm28 & -23\pm14 \\
G211.16$-$19.33North3 & -22\pm5 & \nodata & -22\pm5 \\
G209.94$-$19.52South1 & -22\pm14 & -15\pm27 & -18\pm15 \\
G210.37$-$19.53North & \nodata & -7\pm28 & -7\pm28
\enddata
\tablecomments{Column 1: SCUBA-2 core name,  Column 2-3: Chemical Evolutionary Factor \textbf{(CEF2.0)} for {\ntd} and DNC molecules, respectively, Column 4: \textbf{average CEF2.0}. Three dots indicate no detection.}
\end{deluxetable}

\begin{figure}[th!]
\figurenum{11}
\epsscale{1.1}
\plotone{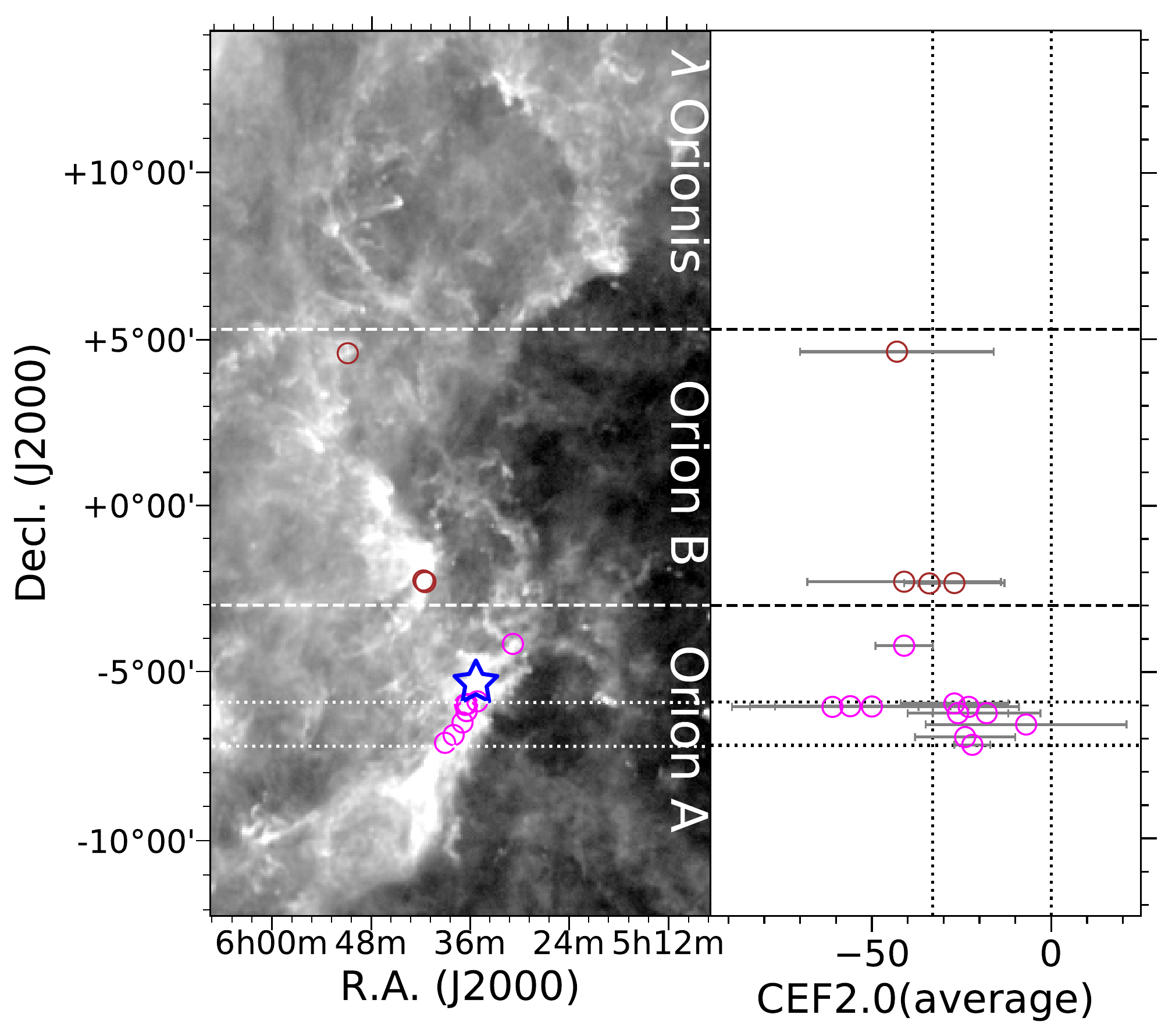}
\caption{Same as Figure \ref{fig:r_ori} but for the spatial distribution and average \textbf{CEF2.0} against declination for \textbf{starless cores} in the Orion region. \textbf{The vertical lines represent CEF2.0 of $-33$ and 0. The horizontal dotted lines represent declination of $-$7\fdg2 and $-$5\fdg9.} \label{fig:cef_ori}}
\end{figure}

\begin{figure}[t!]
\figurenum{12}
\epsscale{1.1}
\plotone{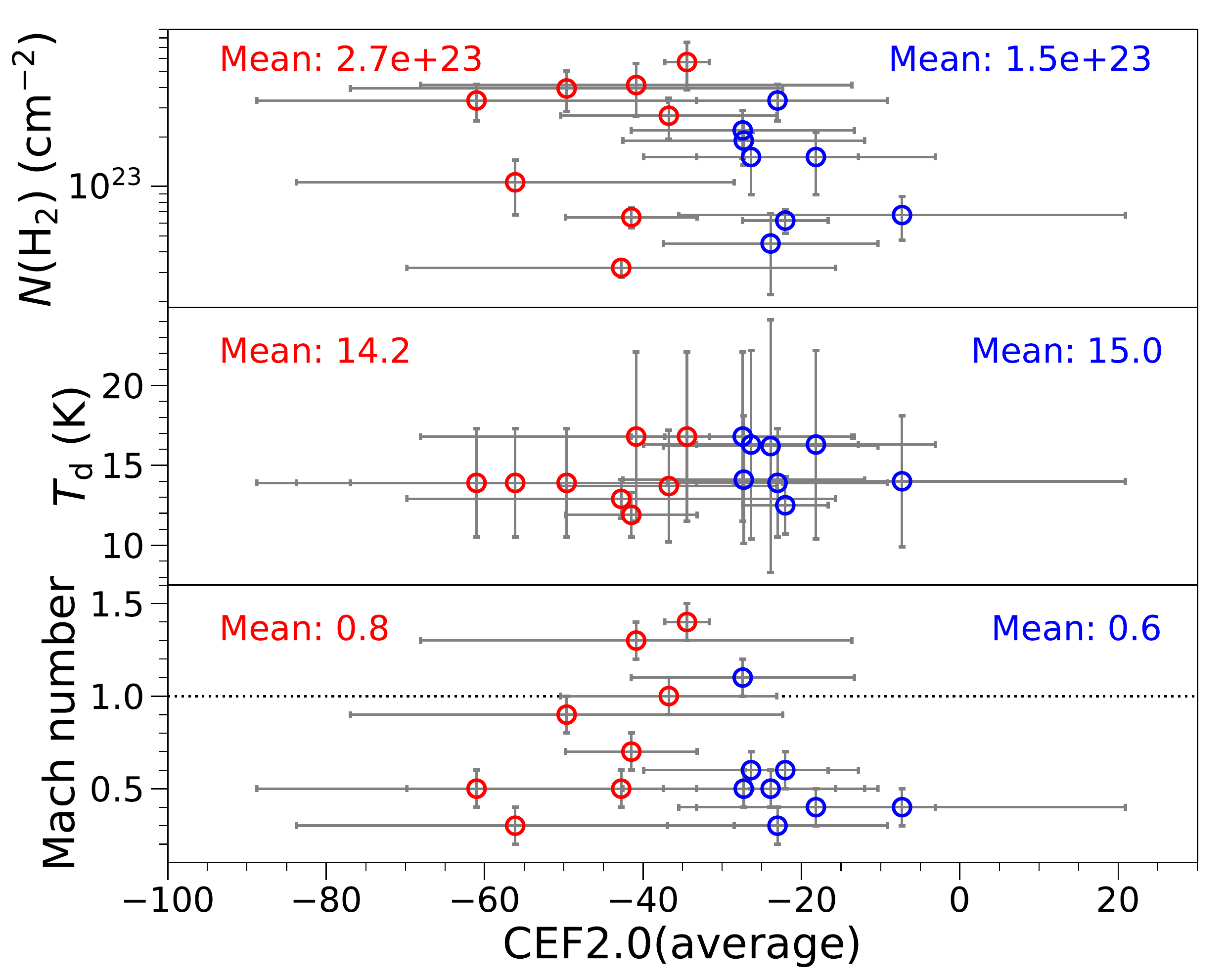}
\caption{H$_2$ column density, dust temperature, and Mach number against average \textbf{CEF2.0} for starless cores in the Orion region. The red and blue represent starless cores with \textbf{average CEF2.0 of ${<}-33$ and ${\geq}-33$, respectively.}  \label{fig:cef_phy}}
\end{figure}

The bottom panel of Figure \ref{fig:cef_ori_sl} shows the \textbf{CEF2.0 of the 16 starless cores in the Orion region and the samples of \citet{Tatematsu:2017hm} aligned in ascending numeric order of the CEF2.0. For the starless Orion cores, the CEF2.0 ranges from ${\sim}-$60 (G209.77$-$19.40East3) to $\sim$0 (G210.37$-$19.53North). Compared to the CEF2.0 for L1544, we judge that at least eight starless cores have CEF2.0 close to zero (${\geq}-33$),} suggesting that they are {\lscore}s on the verge of star formation. Figure \ref{fig:cef_ori} shows their spatial distribution on the \textit{Planck} 850 {\micron} dust continuum map. They are \textbf{mostly located in the south of Orion KL, at Decl. (J2000) = $-$7\fdg2 to $-$5\fdg9.} We suggest that these cores are \textbf{good targets} for studying the initial conditions of star \textbf{formation. G211.16$-$19.33North3, one of these starless cores having CEF2.0 close to zero, and the star-forming Orion core G210.82$-$19.47North1 having bright {\ntd} emission were recently observed by \citet{tatematsu:2020} with the ALMA ACA interferometer.}

\textbf{Regarding the number of the starless cores with successful CEF2.0 estimation, the Orion A sub-region has the largest number. The percentages of the starless cores with successful CEF2.0 estimation are 43{\%} and 40{\%} for Orion A and B, respectively, which are very similar. $\lambda$ Orionis has no core with successful CEF2.0, which is natural because the number of SCUBA-2 cores is small. We see no difference among the Orion sub-regions in terms of the success rate of CEF2.0 estimation.}

We investigate whether there is any variation in the physical properties of starless cores when the average CEF approaches zero. Figure \ref{fig:cef_phy} shows the diagrams for H$_2$ column density, dust temperature, and Mach number as a function of the average \textbf{CEF2.0.} We simply compare the mean values of the physical properties for two groups separated by \textbf{an average CEF2.0 of $-33$: early starless phase (CEF2.0 $<-33$) and late starless phase ($-33\leq$ CEF2.0).} When starless cores evolve, three physical properties do not seem to increase or decrease significantly. The lack of trends in the H$_2$ column density and dust temperature appear to be inconsistent with the prediction of the core evolution model \citep[e.g.,][]{Shirley:2005hy, Aikawa:2008dm}. These discrepancies may be due to the reasons discussed in Section \ref{sec:r_n_dt}. The lack of a trend in the Mach number may suggest that the turbulence dissipation in the dense region of starless core is not required for the onset of star formation, because turbulent dissipation is regarded as one possibility to change a stable core to an unstable one \citep[e.g.,][]{Nakano:1998db}. Other mechanisms may be required to change stable cores to unstable ones \citep[e.g., accretion flow of gas;][]{Gomez:2007ki}. \textbf{Indeed, in G211.16$-$19.33North3, \citet{tatematsu:2020} obtained a hint of gas accretion onto one sub-core inside, with the ALMA ACA interferometer.} However, we cannot rule out the possibility of turbulence dissipation for the beginning of star formation completely because the angular resolution of our telescope may be insufficient, and systematic observations with higher angular resolution are needed.

\subsection{Identification of candidates for {\lscore}s in the environments other than the Orion}
For most of starless cores in the environments other than the Orion region, it is difficult to establish the \textbf{CEF2.0} because of largely different linear beam sizes (Section \ref{sec:r_d}). Moreover, only $N$(DNC)/$N$(\hnc) is available for these environments. Taking this into account, we look for {\lscore} candidates with both high $N$(DNC)/$N$(\hnc) and {\ntd} detection among local ($<$1 kpc) starless cores in two environments. \textbf{All starless cores in the Galactic plane are more distant than 1 kpc. Two cores (SCOPEG001.37$+$20.95, SCOPEG202.32$+$02.53) at high latitudes} are found to have high $N$(DNC)/ $N$(\hnc) and detection in {\ntd} line. Among these cores, the highest $N$(DNC)/$N$(\hnc) is found in SCOPEG001.37$+$ 20.95 (14.6). 

\section{Summary \label{sec:sum}}
We conducted a molecular line survey of 207 SCUBA-2 cores embedded in the Planck Galactic Cold Clumps with the Nobeyama 45-m telescope in the {\ntd} $J = 1-0$, {\nth} $J = 1-0$, DNC $J = 1-0$, {\hnc} $J = 1-0$, CCS $J_N = 7_6-6_5$, CCS $J_N = 8_7-7_6$, {\hcn} $J = 9-8$, and {\cch} $J_{K_a K_c} =  2_{12}-1_{01}$ lines to identify dense cores on the verge of star formation in five different environments ($\lambda$ Orionis, Orion A, B, Galactic plane, and high latitudes). The main results are summarized as follows:
\begin{itemize}
     \item A total of 207 SCUBA-2 cores are classified into 58 starless cores (candidates) and 149 protostellar cores (candidates). They consist of 5 starless cores and 10 protostellar cores in the $\lambda$ Orionis, 24 and 46 in the Orion A, 10 and 18 in the Orion B, 13 and 39 in the Galactic plane, 6 and 36 at high latitudes. Among the total 207 SCUBA-2 cores, starless cores occupy $\sim$28{\%}.
     
     \item The detection rates of early-type molecules (CCS) are low and the detection rates of late-type molecules ({\cch}, {\nth}) and the deuterated molecules (DNC, {\ntd}) are high, suggesting that most of the SCUBA-2 cores are chemically evolved. 
     
     \item \textbf{The integrated-intensity ratio and column-density ratio of {DNC} to {\hnc} (HNC) tend to decrease with increasing distance ($>$1 kpc correspond to linear beam sizes of $>$0.1 pc). This suggests that the deuterium fraction suffers from differential beam dilution between the two lines for distant cores ($>$1 kpc).}
    
     \item For starless and protostellar cores in the $\lambda$ Orionis, Orion A, and B, the column-density ratios of {\ntd}/{\nth}, DNC/{\hnc}, {\nth}/CCS, and {\nth}/{\hcn} are similar. This suggests that cores in these three regions have similar chemical properties. Between four column-density ratios, $N$(DNC)/$N$({\hnc}) and $N$({\ntd})/$N$({\nth}) have a positive correlation, suggesting that the two ratios could act as better tracers for core chemical evolution.

    \item In the Orion region, at least eight starless cores are identified as dense cores on the verge of star formation using the chemical evolution factor \textbf{(CEF2.0)} built on the basis of $N$({\ntd})/$N$({\nth}) and $N$(DNC)/$N$({\hnc}). At high latitudes, at least two starless cores are identified to be close to the beginning of star formation on the basis of high $N$(DNC)/$N$({\hnc}) and {\ntd} line detection. For starless cores in the Orion region, when starless cores evolve, the Mach number does not increase or decrease, which may indicate that the dissipation of turbulence in the dense region of the cores may not be an important mechanism for the onset of star formation as judged from observations with a beam size of 0.04 pc.    
\end{itemize}

SCUBA-2 cores associated with high/low-mass star-forming regions from the Galactic plane to high latitudes are located at highly different distances. For the study of the properties of dense cores at the constant spatial resolution with a single telescope, we need a set of observations to circumvent the beam dilution effect. In addition, for starless cores close to the onset of star formation found in this study, mapping observations are necessary to investigate whether they are gravitationally bound.

\acknowledgments
KT acknowledges Kouji Ohta for discussion. PS is partially supported by a Grant-in-Aid for Scientific Research (KAKENHI Number 18H01259) of Japan Society for the Promotion of Science (JSPS). JHH thanks the National Natural Science Foundation of China under grant Nos. 11873086 and U1631237 and support by the Yunnan Province of China (No.2017HC018). This work is sponsored (in part) by the Chinese Academy of Sciences (CAS), through a grant to the CAS South America Center for Astronomy (CASSACA) in Santiago, Chile. K. W. acknowledges support by the National Key Research and Development Program of China (2017YFA0402702, 2019YFA0405100), the National Science Foundation of China (11973013, 11721303), and the starting grant at the Kavli Institute for Astronomy and Astrophysics, Peking University (7101502287).

\vspace{5mm}
\facilities{No:45m}

\software{\textbf{astropy \citep{2013A&A...558A..33A,2018AJ....156..123A}, GILDAS \citep{2005sf2a.conf..721P, 2013ascl.soft05010G}, NEWSTAR}}


\begin{rotatepage} 
\begin{longrotatetable}
\begin{deluxetable*}{lCCcccCCcl}
\tablecaption{Information of 207 SCUBA-2 cores embedded in Planck Galactic Cold Clumps \label{tbl:target}}
\tablewidth{0pt}
\tabletypesize{\scriptsize}
\tablenum{1}
\tablehead{
\colhead{SCUBA-2 core} & 
\colhead{R.A.(J2000)} & \colhead{Decl.(J2000)}&  
\colhead{YSO ass.} & 
\colhead{Dist.} & 
\colhead{Ref} &
\colhead{$T_{\rm d}$} &
\colhead{$N$(H$_{2}$)} &
\colhead{Env.} &
\colhead{Comment} \\
\colhead{}& \colhead{(hh:mm:ss.ss)} & \colhead{(dd:mm:ss.s)} & \colhead{} & 
\colhead{(kpc)} & \colhead{} & \colhead{(K)} & \colhead{($\times10^{23}$ cm$^{-2}$)} &
\colhead{}  & \colhead{}
}
\decimalcolnumbers
\startdata
G190.15$-$13.75North & 05:19:01.11 & 13:08:05.8 & starless & 0.38 & 11 & 10.8\pm1.7 & 1.1\pm0.2 & OL & PGCC G190.15$-$13.75, $\lambda$ Orionis \\
G190.15$-$13.75South & 05:18:40.55 & 13:01:41.6 & starless & 0.38 & 11 & 10.8\pm1.7 & 1.0\pm0.2 & OL & PGCC G190.15$-$13.75, $\lambda$ Orionis \\
G191.90$-$11.21North & 05:31:28.99 & 12:58:55.0 & starless & 0.38 & 11 & 14.7\pm4.1 & 0.2\pm0.1 & OL & PGCC G191.90$-$11.21, $\lambda$ Orionis \\
G191.90$-$11.21South & 05:31:31.73 & 12:56:15.0 & protostellar & 0.38 & 11 & 14.7\pm4.1 & 0.7\pm0.2 & OL & WISE J053131.58+125615.2, PGCC G191.90$-$11.21, $\lambda$ Orionis \\
G192.12$-$10.90North & 05:33:02.64 & 12:57:53.6 & protostellar & 0.38 & 11 & 13.1\pm1.5 & 0.6\pm0.1 & OL & WISE J053302.57+125810.1, PGCC G192.12$-$10.88, $\lambda$ Orionis \\
G192.12$-$10.90South & 05:32:52.52 & 12:55:08.6 & protostellar & 0.38 & 11 & 13.1\pm1.5 & 0.6\pm0.1 & OL & WISE J053252.62+125509.0, PGCC G192.12$-$10.88, $\lambda$ Orionis \\
G192.12$-$11.10 & 05:32:19.54 & 12:49:40.2 & protostellar & 0.38 & 11 & 13.3\pm3.0 & 1.8\pm0.4 & OL & WISE J053219.43+124942.5, PGCC G192.12$-$11.10, L1582, $\lambda$ Orionis \\
G192.32$-$11.88North & 05:29:54.47 & 12:16:56.0 & protostellar & 0.38 & 11 & 17.3\pm6.0 & 1.1\pm0.4 & OL & [LZK2016] G192N, PGCC G192.32$-$11.88, L1581, $\lambda$ Orionis \\
\enddata
\tablecomments{Column 1: SCUBA-2 core name, Column 2-3: Coordinate in Equatorial system (J2000), Column 4: Young Stellar Object (YSO) association inferred by visual inspection with YSO information of Simbad data and protostar catalogs of \textit{WISE}, \textit{Spitzer}, \textit{Herschel}, and \textit{GAIA}, Column 5: Distance from us, Column 6: Reference for the adopted distance: 1. \citet{Aidelman:2018kd}, 2. \citet{Camargo:2012dm}, 3. \citet{Fischera:2012kc}, 4. \citet{Galli:2018gx}, 5. \citet{Hirota:2008bh}, 6. \citet{Humphreys:1978kq}, 7. \citet{Kounkel:2017eg}, 8. \citet{Lada:2009jb},  9. \citet{Lombardi:2008fk}, 10. \citet{OrtizLeon:2018wz}, 11. \citet{Perryman:1997v}, 12. \citet{Pidopryhora:2015kj}, 13. \citet{Reid:2016gy}, 14. \citet{Straizys:2010fc}, 15. \citet{Straizys:2014kg}, 16. \citet{Straizys:2018fi}, 17. \citet{Sunada:2007fp}, 18. \citet{Wang:2016fa}, Column 7: Dust temperature is from that of PGCC, Column 8: H$_2$ column density is derived from the 850 {\micron} peak intensity of the core and the dust temperature of PGCC \citep{Yi:2018hc, Eden:2019ia}, Column 9: Description of environment surrounding a SCUBA-2 core: `OL', `OA', and `OB' indicate the sub-regions $\lambda$ Orionis, Orion A, and Orion B of the Orion region, respectively, `G' indicates the Galactic plane ($|b|<2\arcdeg$), `H' indicates high latitudes ($|b|\geq2\arcdeg$), Column 10: Source name of YSO, PGCC, or parent cloud. Columns 1-3 and 7-8 are from \citet{Yi:2018hc} and \citet{Eden:2019ia}.}
\end{deluxetable*}
\end{longrotatetable}
\begin{longrotatetable}
\begin{deluxetable*}{lCccCccCccCccccCCC}
\tablecaption{Properties of 82 GHz CCS, 94 GHz CCS, {\hcn}, and {\nth} lines \label{tbl:n2h}}
\tablewidth{0pt}
\tabletypesize{\scriptsize}
\tablenum{3}
\tablehead{
\colhead{} & \multicolumn{3}{c}{82 GHz CCS} & \multicolumn{3}{c}{94 GHz CCS} & \multicolumn{3}{c}{\hcn} & \multicolumn{8}{c}{\nth}
\\
\cmidrule(l{2pt}r{2pt}){2-4}\cmidrule(l{2pt}r{2pt}){5-7}\cmidrule(l{2pt}r{2pt}){8-10}\cmidrule(l{2pt}r{2pt}){11-18}
\colhead{SCUBA-2 core} & \colhead{$T_{\rm peak}$} & \colhead{$V_{\rm LSR}$} & \colhead{$\Delta v$} & \colhead{$T_{\rm peak}$} & \colhead{$V_{\rm LSR}$} & \colhead{$\Delta v$} & \colhead{$T_{\rm peak}$} & \colhead{$V_{\rm LSR}$} & \colhead{$\Delta v$} & \colhead{$T_{\rm peak}$} & \colhead{$V_{\rm LSR, GA}$} & \colhead{$\Delta v_{\rm GA}$} & \colhead{$V_{\rm LSR, HFS}$} & \colhead{$\Delta v_{\rm HFS}$} & \colhead{$T_{\rm ex}$} & \colhead{$\tau$} & \colhead{$T_{\rm ant}\tau$}\\
\cmidrule(l{2pt}r{2pt}){3-4}\cmidrule(l{2pt}r{2pt}){6-7}\cmidrule(l{2pt}r{2pt}){9-10}\cmidrule(l{2pt}r{2pt}){12-15}
\colhead{} & \colhead{(K)} & \multicolumn{2}{c}{(km s$^{-1}$)} & \colhead{(K)} & \multicolumn{2}{c}{(km s$^{-1}$)} & \colhead{(K)} & \multicolumn{2}{c}{(km s$^{-1}$)} & \colhead{(K)} & \multicolumn{4}{c}{(km s$^{-1}$)} & \colhead{(K)} & \colhead{}   & \colhead{(K)} 
}
\decimalcolnumbers
\startdata
G190.15$-$13.75North & \leq0.31 & \nodata & \nodata & \leq0.29 & \nodata & \nodata & \leq0.27 & \nodata & \nodata & \leq0.29 & \nodata & \nodata & \nodata & \nodata & \nodata & \nodata & \nodata \\
G190.15$-$13.75South & \leq0.30 & \nodata & \nodata & \leq0.29 & \nodata & \nodata & \leq0.28 & \nodata & \nodata & 0.51 & 1.34 & 0.19 & 1.35 & 0.30 & 5.4\pm2.7^{\dagger} & \nodata & 2.8\pm0.4 \\
G191.90$-$11.21North & \leq0.29 & \nodata & \nodata & 0.37 & 5.62 & 0.24 & \leq0.25 & \nodata & \nodata & \leq0.26 & \nodata & \nodata & \nodata & \nodata & \nodata & \nodata & \nodata \\
G191.90$-$11.21South & \leq0.26 & \nodata & \nodata & \leq0.24 & \nodata & \nodata & \leq0.23 & \nodata & \nodata & 0.72 & 10.53 & 0.38 & 10.50 & 0.40 & 4.2\pm0.3 & 7.6\pm2.8 & 10.9\pm2.1 \\
G192.12$-$10.90North & 0.40 & 10.17 & 0.34 & 0.42 & 10.07 & 0.15 & 0.56 & 10.08 & 0.49 & 0.94 & 10.04 & 0.33 & 10.00 & 0.41 & 7.0\pm0.7 & 1.8\pm1.8 & 7.6\pm1.3 \\
G192.12$-$10.90South & 0.30 & 10.28 & 0.10 & \leq0.27 & \nodata & \nodata & \leq0.21 & \nodata & \nodata & \leq0.26 & \nodata & \nodata & \nodata & \nodata & \nodata & \nodata & \nodata \\
G192.12$-$11.10 & \leq0.24 & \nodata & \nodata & \leq0.24 & \nodata & \nodata & \leq0.22 & \nodata & \nodata & 0.52 & 10.02 & 0.35 & 10.10 & 0.73 & 4.2\pm0.3 & 3.0\pm1.8 & 4.5\pm0.8 \\
G192.32$-$11.88North & \leq0.29 & \nodata & \nodata & \leq0.24 & \nodata & \nodata & \leq0.25 & \nodata & \nodata & 0.85 & 12.12 & 0.59 & 12.20 & 0.66 & 4.3\pm0.2 & 7.3\pm1.8 & 11.2\pm1.6 \\
\enddata
\tablecomments{All values are measured in a spectrum whose peak temperature is higher than 3$\sigma$. In the case of a peak temperature below 3$\sigma$, the 3$\sigma$ level is listed as an upper limit. Column 1: SCUBA-2 core name, Column 2-4: Peak temperature at the $T_{\rm A}^{\ast}$ scale, systemic velocity, Full Width at Half Maximum (FWHM) inferred by Gaussian fitting to the spectrum of 82 GHz CCS line, respectively, Column 5-7: Same as Column 2-4 but of 94 GHz CCS line, Column 8-10: Same as Column 2-4 but of {\hcn}, Column 11-13: Peak temperature at the $T_{\rm A}^{\ast}$ scale, systemic velocity, and FWHM  inferred by Gaussian fitting to the brightest hyperfine component of {\nth} line, Column 14-18: Systemic velocity, FWHM, excitation temperature ($T_{\rm ex}$), \textbf{total line optical depth of all the hyperfine components ($\tau$), and $T_{\rm ant}\tau~(= (T_{\rm ex} - T_{\rm bg})\tau$) estimated through the hyperfine structure fitting to seven components of the {\nth} line. In columns 16 and 17, `${\dagger}$' indicates that the value is derived from dust temperature.}}
\end{deluxetable*}
\end{longrotatetable}
\begin{longrotatetable}
\begin{deluxetable*}{lCccCccCccCccccCCC}
\tablecaption{Properties of {\cch}, DNC, {\hnc}, and {\ntd} lines \label{tbl:n2d}}
\tablewidth{0pt}
\tabletypesize{\scriptsize}
\tablenum{4}
\tablehead{
\colhead{} & \multicolumn{3}{c}{\cch} & \multicolumn{3}{c}{DNC} & \multicolumn{3}{c}{\hnc} & \multicolumn{8}{c}{\ntd}
\\
\cmidrule(l{2pt}r{2pt}){2-4}\cmidrule(l{2pt}r{2pt}){5-7}\cmidrule(l{2pt}r{2pt}){8-10}\cmidrule(l{2pt}r{2pt}){11-18}
\colhead{SCUBA-2 core} & \colhead{$T_{\rm peak}$} & \colhead{$V_{\rm LSR}$} & \colhead{$\Delta v$} & \colhead{$T_{\rm peak}$} & \colhead{$V_{\rm LSR}$} & \colhead{$\Delta v$} & \colhead{$T_{\rm peak}$} & \colhead{$V_{\rm LSR}$} & \colhead{$\Delta v$} & \colhead{$T_{\rm peak}$} & \colhead{$V_{\rm LSR, GA}$} & \colhead{$\Delta v_{\rm GA}$} & \colhead{$V_{\rm LSR, HFS}$} & \colhead{$\Delta v_{\rm HFS}$} & \colhead{$T_{\rm ex}$} & \colhead{$\tau$} & \colhead{$T_{\rm ant}\tau$}\\
\cmidrule(l{2pt}r{2pt}){3-4}\cmidrule(l{2pt}r{2pt}){6-7}\cmidrule(l{2pt}r{2pt}){9-10}\cmidrule(l{2pt}r{2pt}){12-15}
\colhead{} & \colhead{(K)} & \multicolumn{2}{c}{(km s$^{-1}$)} & \colhead{(K)} & \multicolumn{2}{c}{(km s$^{-1}$)} & \colhead{(K)} & \multicolumn{2}{c}{(km s$^{-1}$)} &  \colhead{(K)} & \multicolumn{4}{c}{(km s$^{-1}$)} & \colhead{(K)} & \colhead{}   & \colhead{(K)} 
}
\decimalcolnumbers
\startdata
G190.15$-$13.75North & \leq0.13 & \nodata & \nodata & \leq0.20 & \nodata & \nodata & \leq0.15 & \nodata & \nodata & \leq0.21 & \nodata & \nodata & \nodata & \nodata & \nodata & \nodata & \nodata \\
G190.15$-$13.75South & \leq0.13 & \nodata & \nodata & \leq0.18 & \nodata & \nodata & \leq0.14 & \nodata & \nodata & \leq0.18 & \nodata & \nodata & \nodata & \nodata & \nodata & \nodata & \nodata \\
G191.90$-$11.21North & \leq0.12 & \nodata & \nodata & \leq0.16 & \nodata & \nodata & \leq0.13 & \nodata & \nodata & \leq0.14 & \nodata & \nodata & \nodata & \nodata & \nodata & \nodata & \nodata \\
G191.90$-$11.21South & 0.38 & 10.49 & 0.49 & 0.44 & 10.59 & 1.00 & 0.11 & 10.52 & 0.76 & 0.16 & 10.73 & 0.26 & \nodata & \nodata & 7.4\pm3.7^{\dagger} & \nodata & \nodata \\
G192.12$-$10.90North & 0.77 & 10.02 & 0.59 & 0.52 & 10.15 & 0.89 & 0.18 & 9.99 & 0.99 & 0.19 & 10.01 & 0.35 & \nodata & \nodata & 6.5\pm3.2^{\dagger} & \nodata & \nodata \\
G192.12$-$10.90South & 0.21 & 10.11 & 0.54 & \leq0.16 & \nodata & \nodata & \leq0.10 & \nodata & \nodata & \leq0.13 & \nodata & \nodata & \nodata & \nodata & \nodata & \nodata & \nodata \\
G192.12$-$11.10 & 0.37 & 10.12 & 0.85 & 0.33 & 10.32 & 1.35 & 0.13 & 10.43 & 1.09 & \leq0.13 & \nodata & \nodata & \nodata & \nodata & \nodata & \nodata & \nodata \\
G192.32$-$11.88North & 0.47 & 12.14 & 0.64 & 0.69 & 12.13 & 1.02 & 0.23 & 12.31 & 0.60 & 0.18 & 12.02 & 0.68 & \nodata & \nodata & 8.7\pm4.3^{\dagger} & \nodata & \nodata \\
\enddata
\tablecomments{All values are measured in a spectrum whose peak temperature is higher than 3$\sigma$. In the case of a peak temperature below 3$\sigma$, the 3$\sigma$ level is listed as an upper limit. Column 1: SCUBA-2 core name, Column 2-4: Peak temperature at the $T_{\rm A}^{\ast}$ scale, systemic velocity, Full Width at Half Maximum (FWHM) inferred by Gaussian fitting to the spectrum of {\cch} line, respectively, Column 5-7: Same as Column 2-4 but of DNC, Column 8-10: Same as Column 2-4 but of {\hnc}, Column 11-13: Peak temperature at the $T_{\rm A}^{\ast}$ scale, systemic velocity, and FWHM inferred by Gaussian fitting to the brightest hyperfine component of {\ntd} line, Column 14-18: Systemic velocity, FWHM, excitation temperature ($T_{\rm ex}$), \textbf{total line optical depth of all the hyperfine components ($\tau$), and $T_{\rm ant}\tau~(= (T_{\rm ex} - T_{\rm bg})\tau$) estimated through the hyperfine structure fitting to seven components of the {\ntd} line. In columns 16 and 17, `${\dagger}$' indicates that the value is derived from dust temperature.}}
\end{deluxetable*}
\end{longrotatetable}
\begin{longrotatetable}
\begin{deluxetable*}{lCCCCCCCCCCCC}
\tablecaption{Column density, column-density ratio, Mach number, and integrated intensity ratio of {\ntd}, {\nth}, DNC, {\hnc}, CCS, and {\hcn} molecules \label{tbl:n_r}}
\tablewidth{0pt}
\tabletypesize{\scriptsize}
\tablenum{5}
\tablehead{
\colhead{SCUBA-2 core} & \colhead{$N$(\ntd)} & \colhead{$N$(\nth)} & \colhead{$N$(DNC)} & \colhead{$N$(\hnc)} &  \colhead{$N$(CCS)} & \colhead{$N$(\hcn)}  &  \colhead{$\frac{{N({\rm N_2D^+})}}{{N({\rm N_2H^+})}}$} & \colhead{$\frac{N({\rm DNC})}{N({\rm HN^{13}C})}$} & \colhead{$\frac{N({\rm N_2H^+})}{N({\rm CCS})}$} & \colhead{$\frac{N({\rm N_2H^+})}{N({\rm HC_3N})}$} & \colhead{$M$} & \colhead{$\frac{\int T({\rm DNC})dv}{\int T({\rm HN^{13}C})dv}$} \\
\colhead{} & \colhead{($10^{11}$ cm$^{-2}$)} & \colhead{($10^{12}$ cm$^{-2}$)} & \colhead{($10^{11}$ cm$^{-2}$)} & \colhead{($10^{11}$ cm$^{-2}$)} & \colhead{($10^{11}$ cm$^{-2}$)} & \colhead{($10^{11}$ cm$^{-2}$)} & \colhead{} & \colhead{} & \colhead{} & \colhead{} & \colhead{} 
}
\decimalcolnumbers
\startdata
G190.15$-$13.75North & \leq15.0 & \leq1.8 & \leq4.8 & \leq3.8 & \leq53.0 & \leq62.0 & \nodata & \nodata & \nodata & \nodata & \nodata & \nodata \\
G190.15$-$13.75South & \leq13.0 & 1.0\pm0.1^{\dagger} & \leq4.2 & \leq3.5 & \leq51.0 & \leq65.0 & \leq1.30 & \nodata & \geq0.2 & \geq0.2 & 0.6\pm0.1 & \nodata \\
G191.90$-$11.21North & \leq9.4 & \leq1.4 & \leq3.4 & \leq2.8 & \leq25.0 & \leq24.0 & \nodata & \nodata & \nodata & \nodata & \nodata & \nodata \\
G191.90$-$11.21South & 5.5\pm3.0^{\dagger} & 6.9\pm1.3 & 20.0\pm10.0 & 3.5\pm1.8 & \leq23.0 & \leq22.0 & 0.08\pm0.05 & 5.7\pm4.1 & \geq3.0 & \geq3.1 & 0.7\pm0.1 & 4.5\pm0.5 \\
G192.12$-$10.90North & 8.8\pm4.6^{\dagger} & 3.7\pm0.6 & 22.0\pm11.0 & 8.1\pm4.1 & 30.0\pm16.0 & 79.0\pm40.0 & 0.24\pm0.13 & 2.7\pm1.9 & 1.2\pm0.7 & 0.5\pm0.2 & 0.8\pm0.1 & 2.1\pm0.2 \\
G192.12$-$10.90South & \leq8.5 & \leq1.4 & \leq3.4 & \leq2.3 & 6.6\pm4.2 & \leq28.0 & \nodata & \nodata & \leq2.1 & \nodata & \nodata & \nodata \\
G192.12$-$11.10 & \leq8.5 & 5.1\pm0.9 & 20.0\pm10.0 & 6.3\pm3.2 & \leq25.0 & \leq27.0 & \leq0.17 & 3.2\pm2.3 & \geq2.0 & \geq1.9 & 1.4\pm0.2 & 2.7\pm0.4 \\
G192.32$-$11.88North & 17.0\pm8.7^{\dagger} & 11.0\pm1.6 & 34.0\pm17.0 & 6.5\pm3.3 & \leq21.0 & \leq18.0 & 0.15\pm0.08 & 5.2\pm3.7 & \geq5.2 & \geq6.1 & 1.1\pm0.2 & 4.0\pm0.4 \\
\enddata
\tablecomments{All values are listed for the source whose peak temperature is higher than 3$\sigma$. Column 1: SCUBA-2 core name, Column 2-7: Column densities of {\ntd}, {\nth}, DNC, {\hnc}, 82 GHz CCS, and {\hcn}, Column 8-11: Ratio of the column densities, Column 12: Mach number of {\nth} line\textbf{, and Column 13: Integrated intensity ratio of DNC to {\hnc} lines. `${\dagger}$' indicates that the column density is derived from dust temperature and `${\ddagger}$' indicates the lower limit when the assumed $T_{\rm ex}$ is too low to constrain the upper bound.}}
\end{deluxetable*}
\end{longrotatetable}
\begin{rotatetable*}
\begin{deluxetable*}{clcccccccccccccccc}
\tablecaption{Minimum, maximum, mean, and median of column-density ratios of {\ntd}/{\nth}, DNC/{\hnc}, {\nth}/CCS, and {\nth}/{\hcn} for 207 SCUBA-2 cores in 
different environments \label{tbl:r_stat}}
\tablewidth{0pt}
\tabletypesize{\scriptsize}
\tablenum{7}
\tablehead{
\colhead{Environment} & \colhead{Source} & \multicolumn{4}{c}{$\frac{{\rm N(N_2D^+)}}{{\rm N(N_2H^+)}}$} & \multicolumn{4}{c}{$\frac{\rm N(DNC)}{\rm N(HN^{13}C)}$} & \multicolumn{4}{c}{$\frac{\rm N(N_2H^+)}{\rm N(CCS)}$} & \multicolumn{4}{c}{$\frac{\rm N(N_2H^+)}{\rm N(HC_3N)}$}  \\
\cmidrule(l{2pt}r{2pt}){3-6}\cmidrule(l{2pt}r{2pt}){7-10}\cmidrule(l{2pt}r{2pt}){11-14}\cmidrule(l{2pt}r{2pt}){15-18}
& & \colhead{min.} & \colhead{max.} & \colhead{mean} & \colhead{median} & \colhead{min.} & \colhead{max.} & \colhead{mean} & \colhead{median} &  \colhead{min.} & \colhead{max.} & \colhead{mean} & \colhead{median} &  \colhead{min.} & \colhead{max.} & \colhead{mean} & \colhead{median} 
}
\startdata
\multirow{3}{*}{Orion region} & Both  & 0.04 & 0.4 & 0.2 & 0.2 & 0.7 & 7.7 & 3.3 & 3.0 & 0.8 & 20.8 & 8.7 & 7.0 & 0.1 & 11.2 & 2.7 & 2.1 \\
 & Starless core & 0.06 & 0.3 & 0.2 & 0.2 & 0.7 & 7.7 & 4.0 & 4.0 & 7.6 & 7.6 & 7.6 & 7.6 & 0.6 & 8.2 & 3.2 & 2.4 \\
 & Protostellar core & 0.04 & 0.4 & 0.2 & 0.1 & 1.0 & 5.7 & 2.9 & 2.7 & 0.8 & 20.8 & 8.9 & 5.3 & 0.1 & 11.2 & 2.4 & 1.5 \\
\hline
\multirow{3}{*}{$\lambda$ Orionis} & Both  & 0.08 & 0.2 & 0.1 & 0.2 & 1.8 & 5.7 & 4.0 & 3.8 & 1.2 & 1.2 & 1.2 & 1.2 & 0.5 & 2.6 & 1.3 & 1.4 \\
 & Starless core & \nodata & \nodata & \nodata & \nodata & \nodata & \nodata & \nodata & \nodata & \nodata & \nodata & \nodata & \nodata & \nodata & \nodata & \nodata & \nodata \\
 & Protostellar core & 0.08 & 0.2 & 0.1 & 0.2 & 1.8 & 5.7 & 4.0 & 3.8 & 1.2 & 1.2 & 1.2 & 1.2 & 0.5 & 2.6 & 1.3 & 1.4 \\
\hline
\multirow{3}{*}{Orion A} & Both  & 0.05 & 0.4 & 0.2 & 0.1 & 0.7 & 7.7 & 3.4 & 2.9 & 0.8 & 20.8 & 9.7 & 7.3 & 0.1 & 11.2 & 2.9 & 2.2 \\
 & Starless core & 0.06 & 0.3 & 0.2 & 0.2 & 0.7 & 7.7 & 4.0 & 4.1 & 7.6 & 7.6 & 7.6 & 7.6 & 1.1 & 4.7 & 2.8 & 2.2 \\
\enddata
\tablecomments{Only cores where both column densities are successfully estimated are considered. The {\nth}, CCS, and {\hcn} data are available for only the Orion region.}
\end{deluxetable*}
\end{rotatetable*}
\end{rotatepage}
\begin{figure*}
\epsscale{1}
\figurenum{3}
\plotone{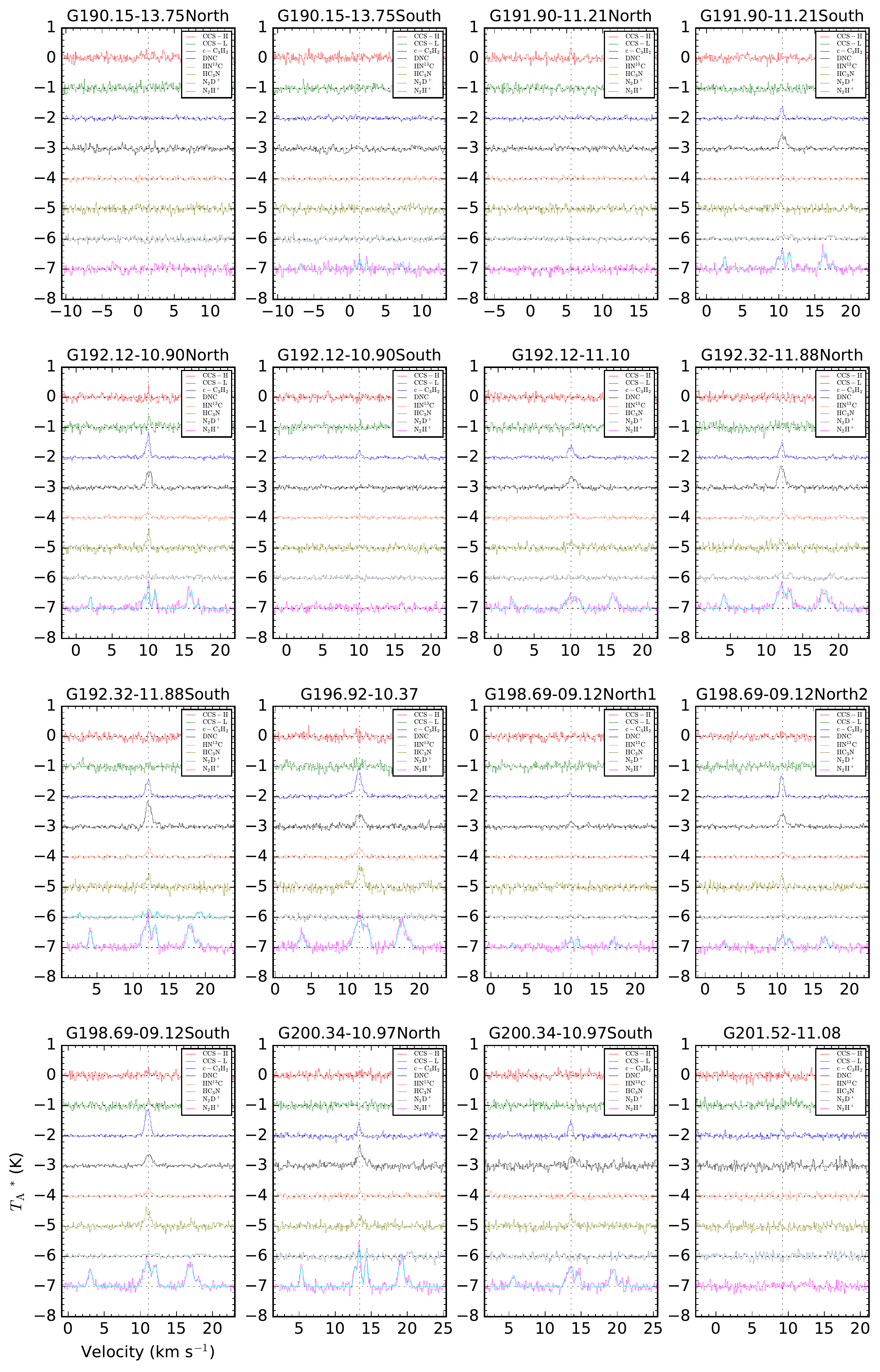}
\caption{Spectra of 207 SCUBA-2 cores in Planck Galactic Cold Clump cores in eight molecular lines. A vertical line indicates a systemic velocity. A horizontal line represents 0 K in {\ta} scale. \textbf{For the {\nth} and {\ntd} emission, the cyan line represents the hyperfine structure fitting to result.} \label{fig:spt}}
\end{figure*}

\begin{figure*}
\figurenum{3}
\epsscale{1}         
\plotone{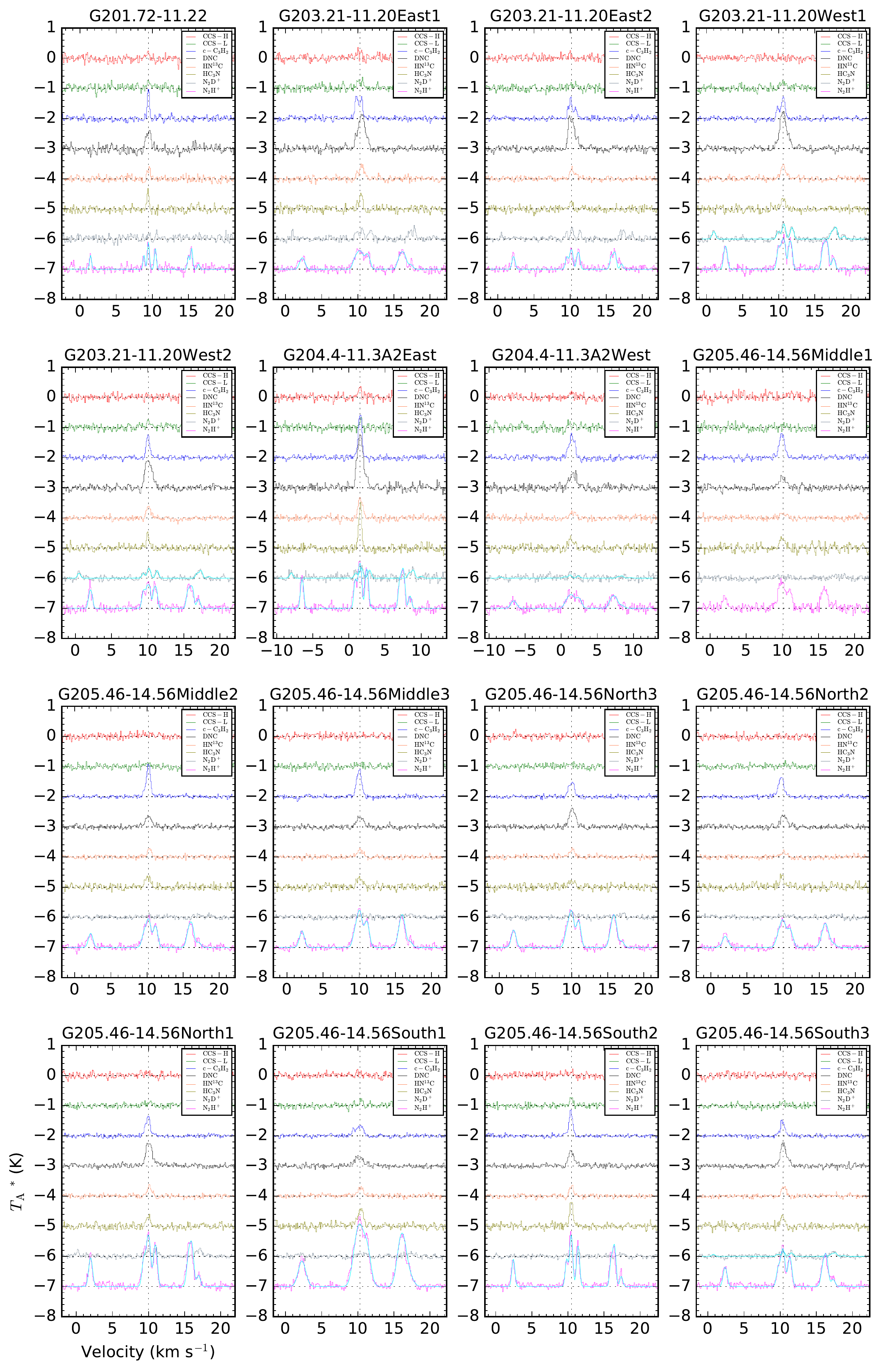}
\caption{(continued)}
\end{figure*}

\begin{figure*}
\figurenum{3}
\epsscale{1}
\plotone{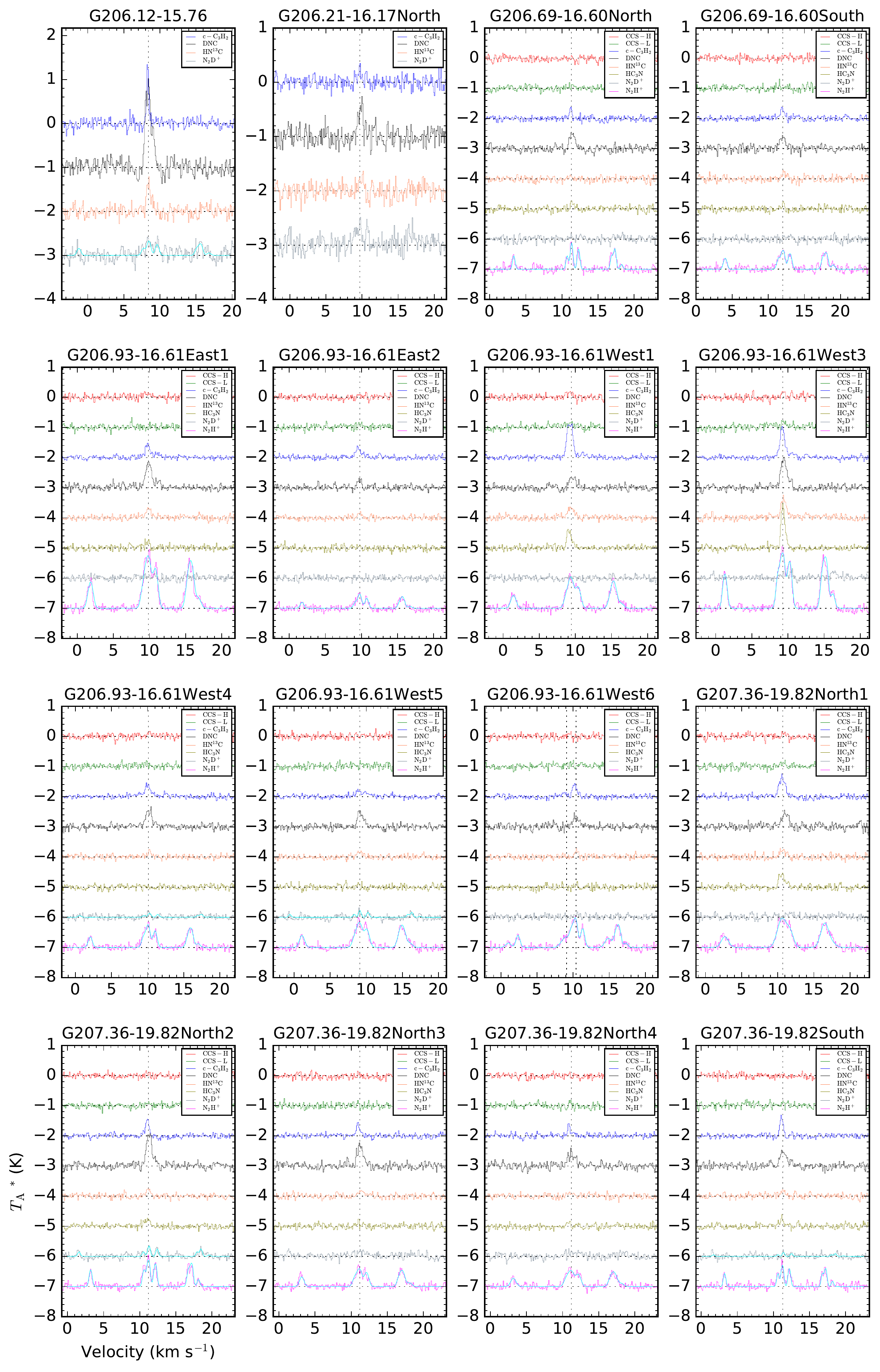}
\caption{(continued)}
\end{figure*}

\begin{figure*}
\figurenum{3}
\epsscale{1}
\plotone{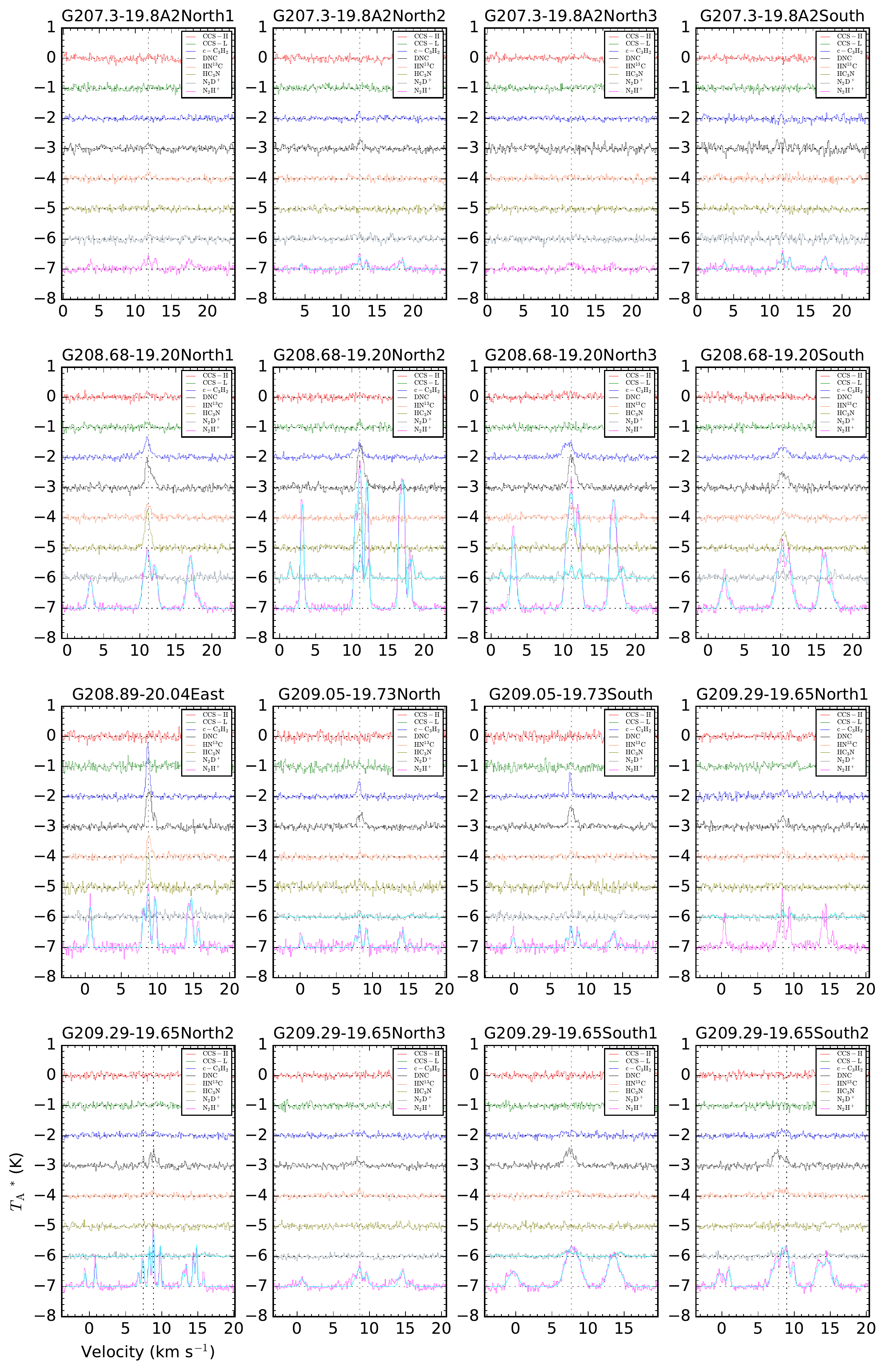}
\caption{(continued)}
\end{figure*}

\begin{figure*}
\figurenum{3}
\epsscale{1}
\plotone{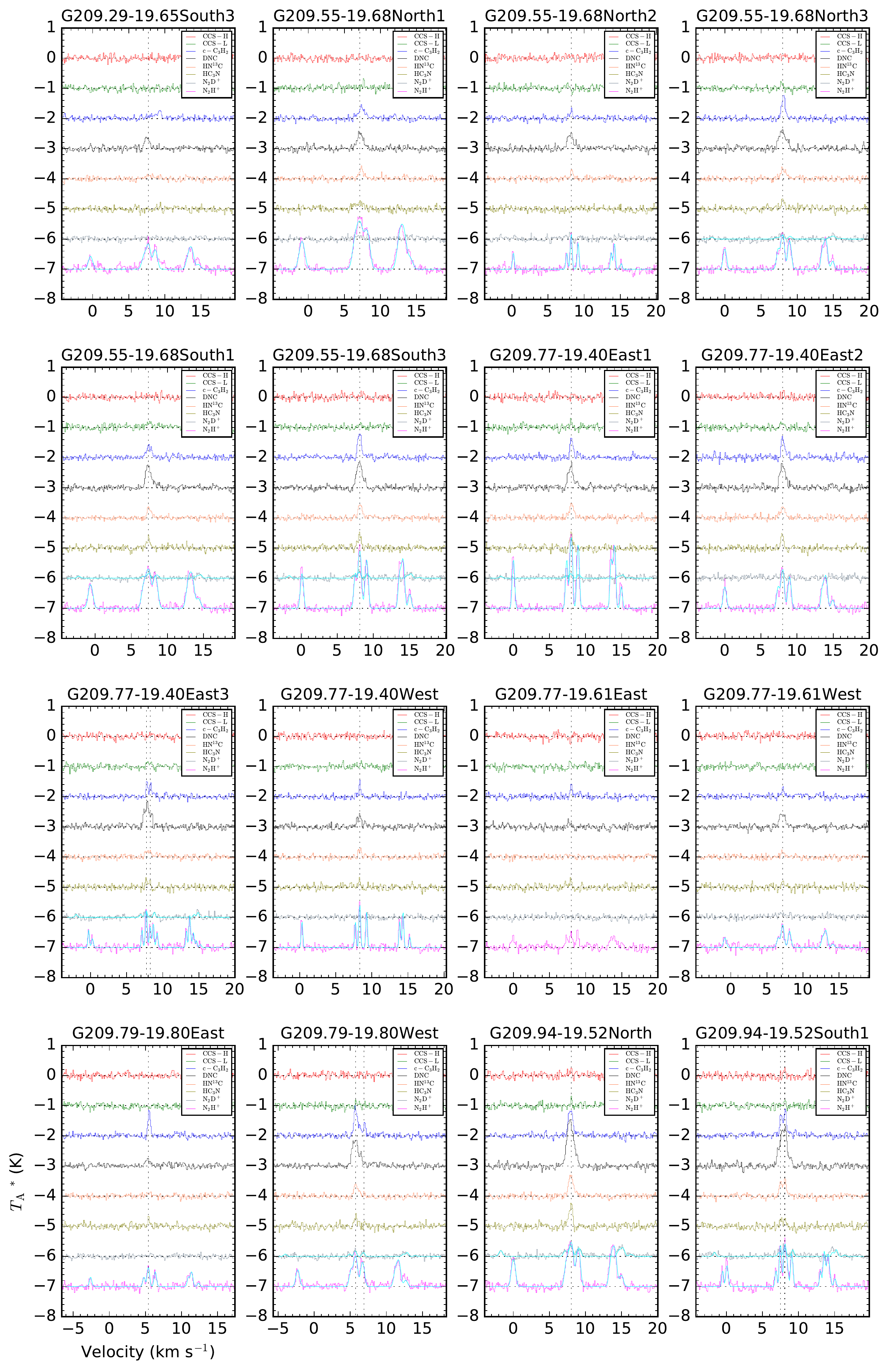}
\caption{(continued)}
\end{figure*}

\begin{figure*}
\figurenum{3}
\epsscale{1}
\plotone{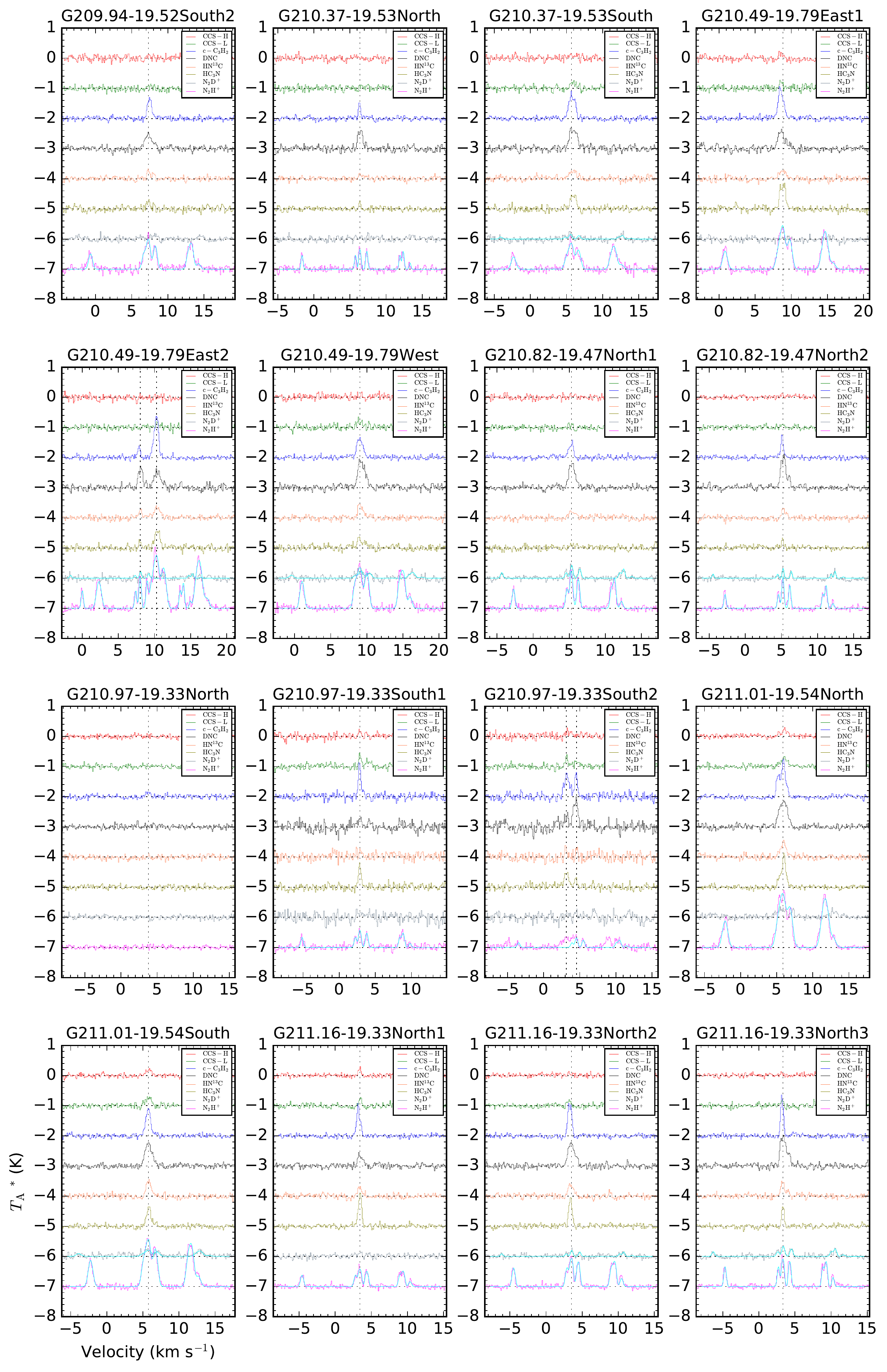}
\caption{(continued)}
\end{figure*}

\begin{figure*}
\figurenum{3}
\epsscale{1}
\plotone{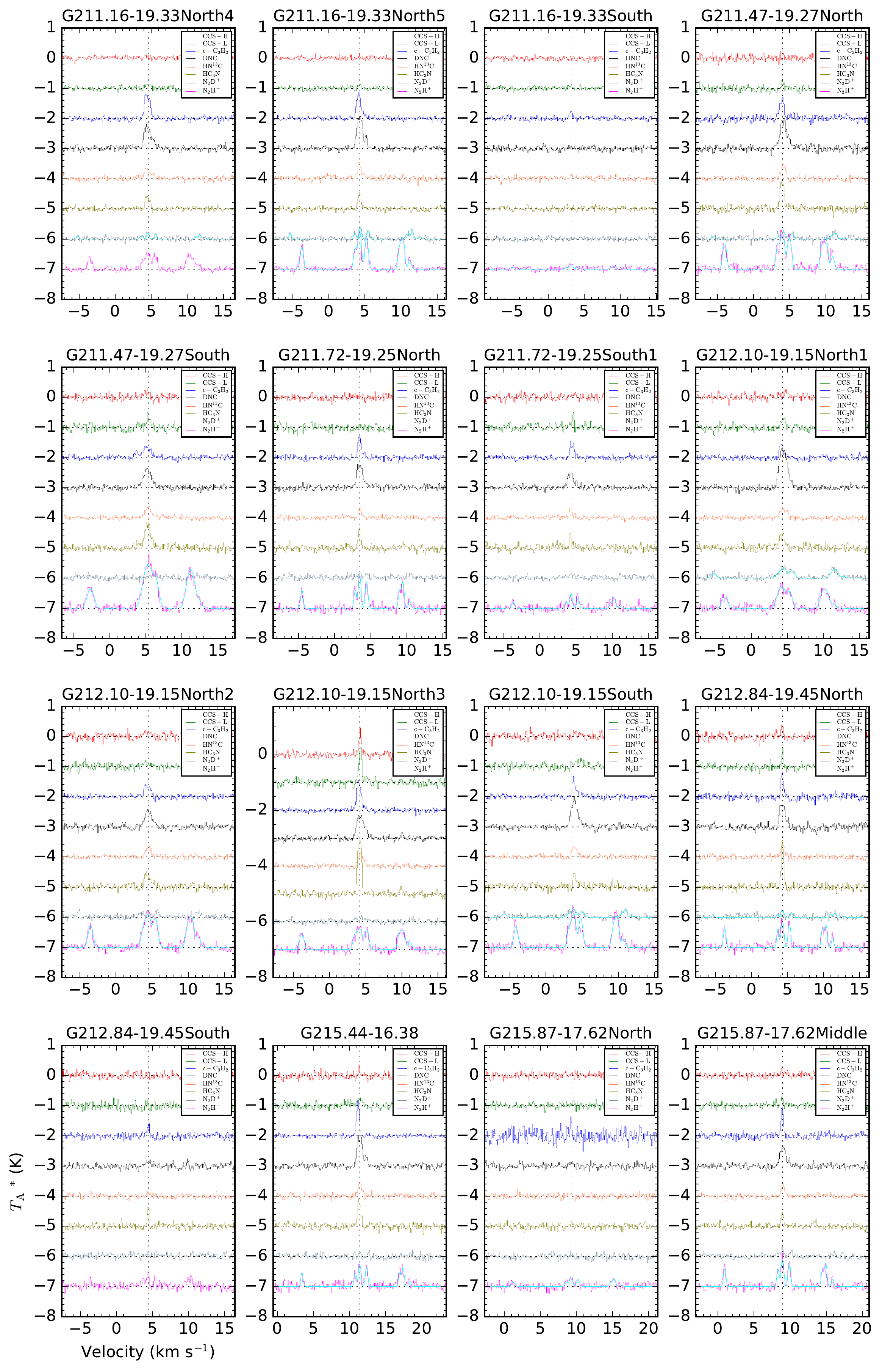}
\caption{(continued)}
\end{figure*}

\begin{figure*}
\figurenum{3}
\epsscale{1}
\plotone{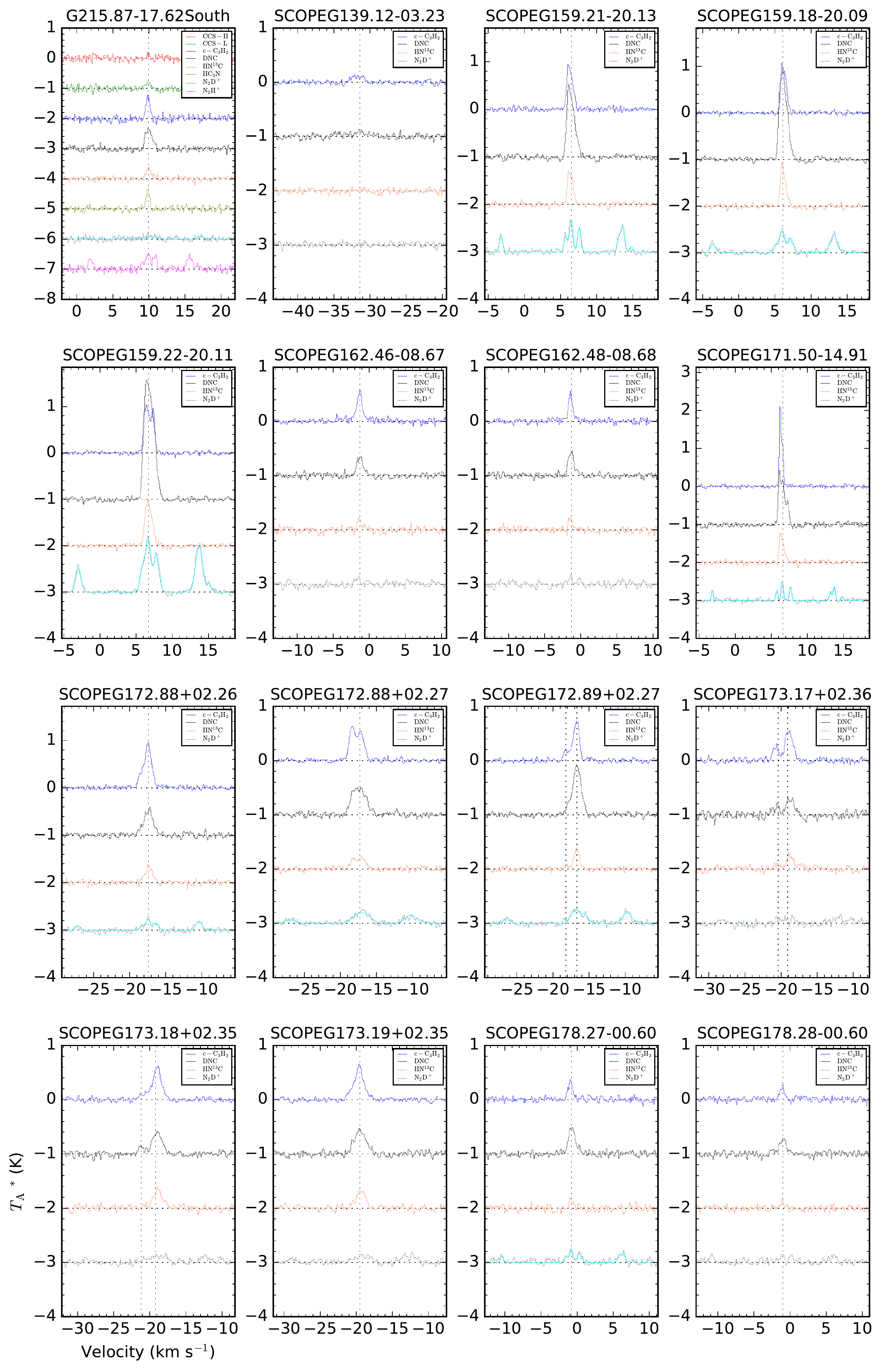}
\caption{(continued)}
\end{figure*}

\begin{figure*}
\figurenum{3}
\epsscale{1}
\plotone{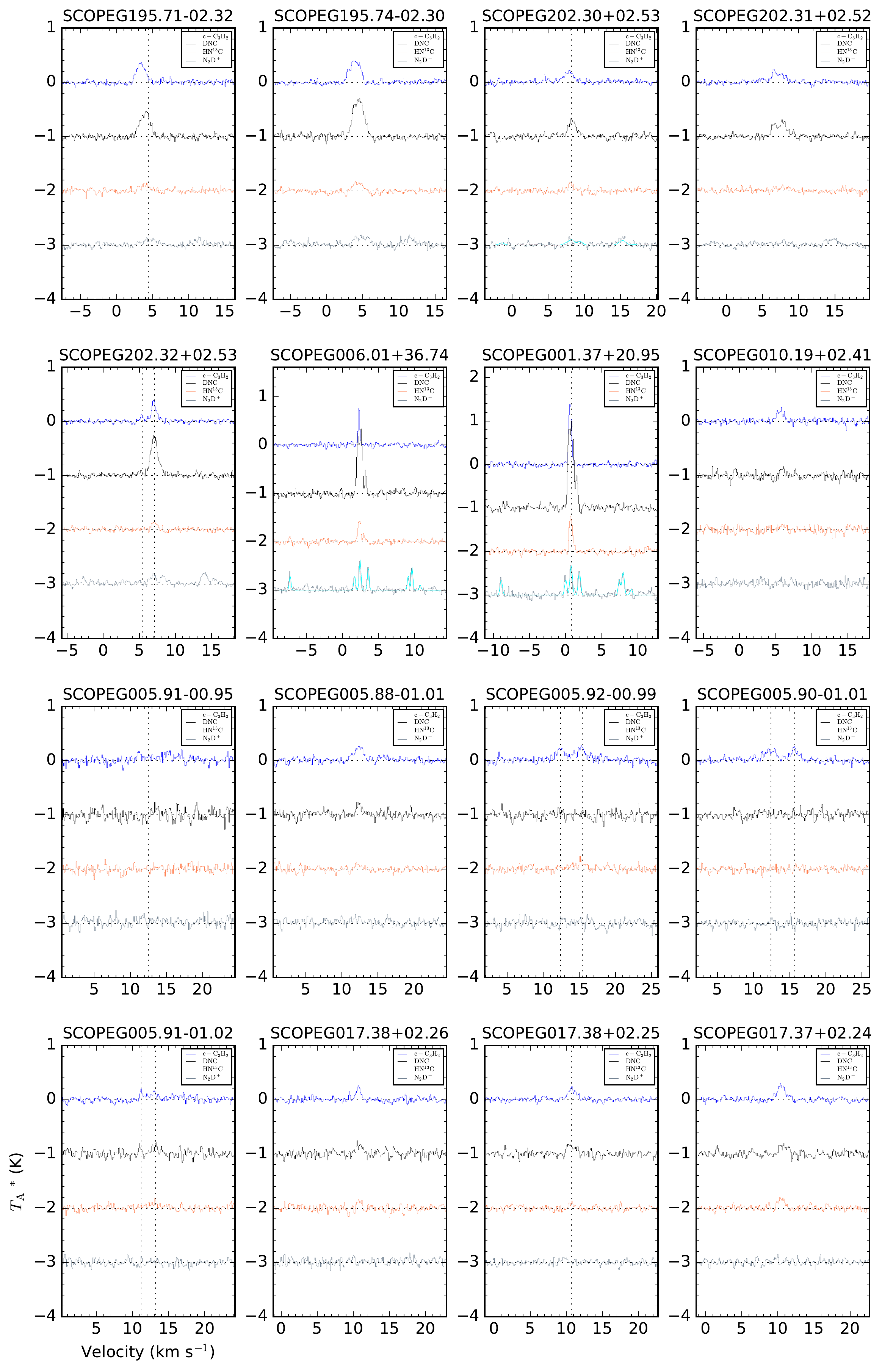}
\caption{(continued)}
\end{figure*}

\begin{figure*}[htb!]
\figurenum{3}
\plotone{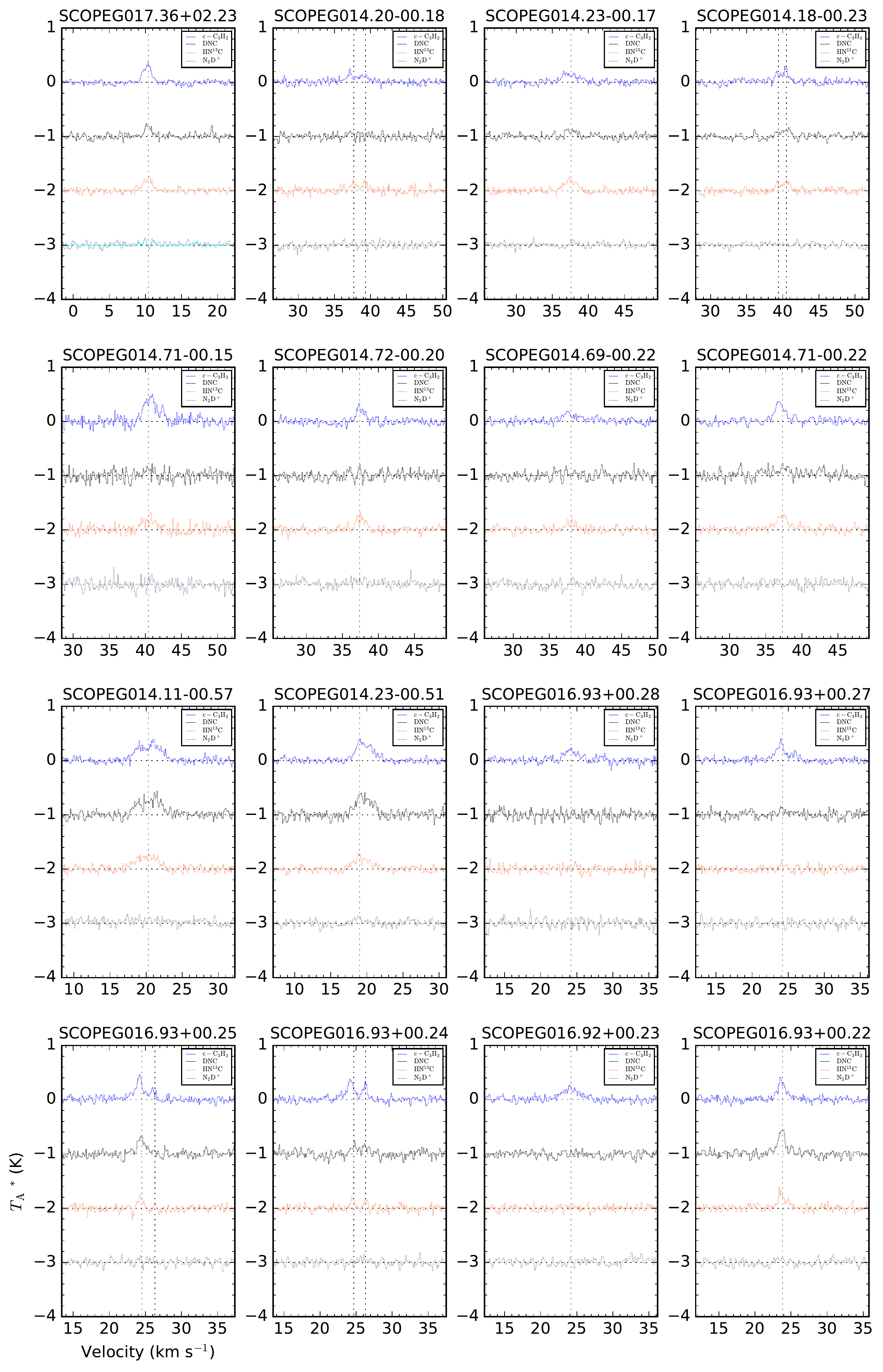}
\caption{(continued)}
\end{figure*}

\begin{figure*}
\figurenum{3}
\epsscale{1}
\plotone{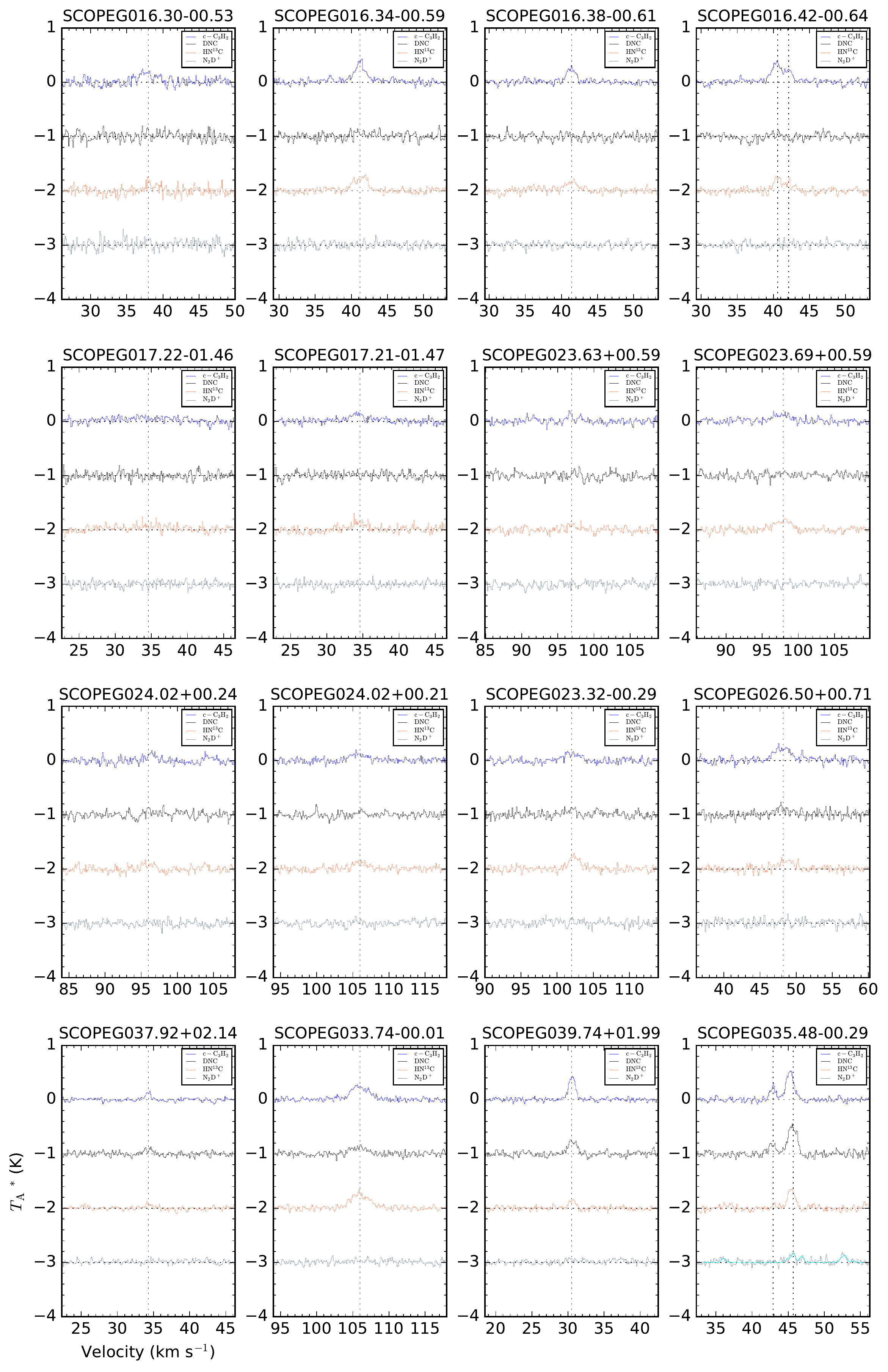}
\caption{(continued)}
\end{figure*}

\begin{figure*}
\figurenum{3}
\epsscale{1}
\plotone{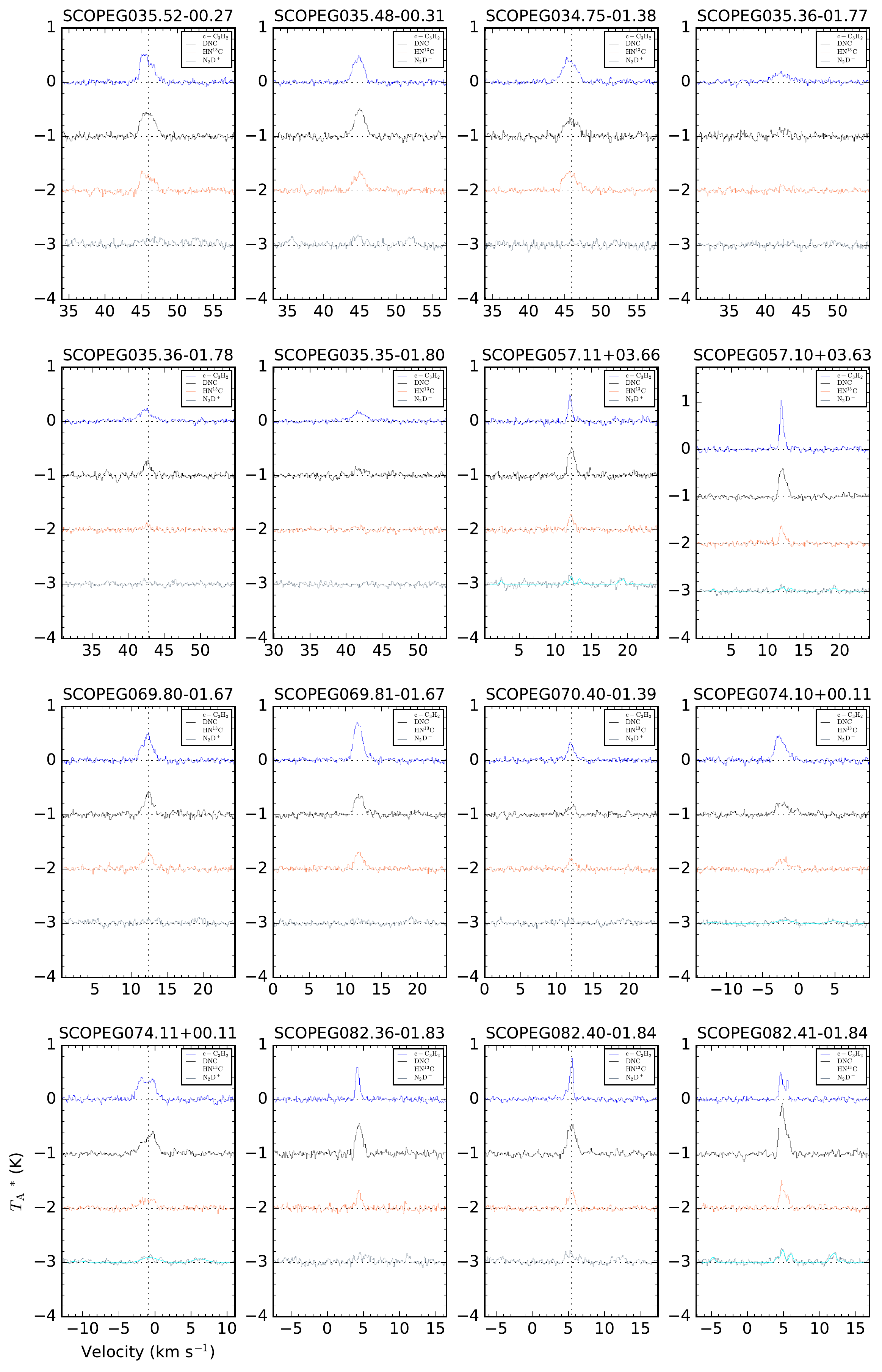}
\caption{(continued)}
\end{figure*}

\begin{figure*}
\figurenum{3}
\epsscale{1}
\plotone{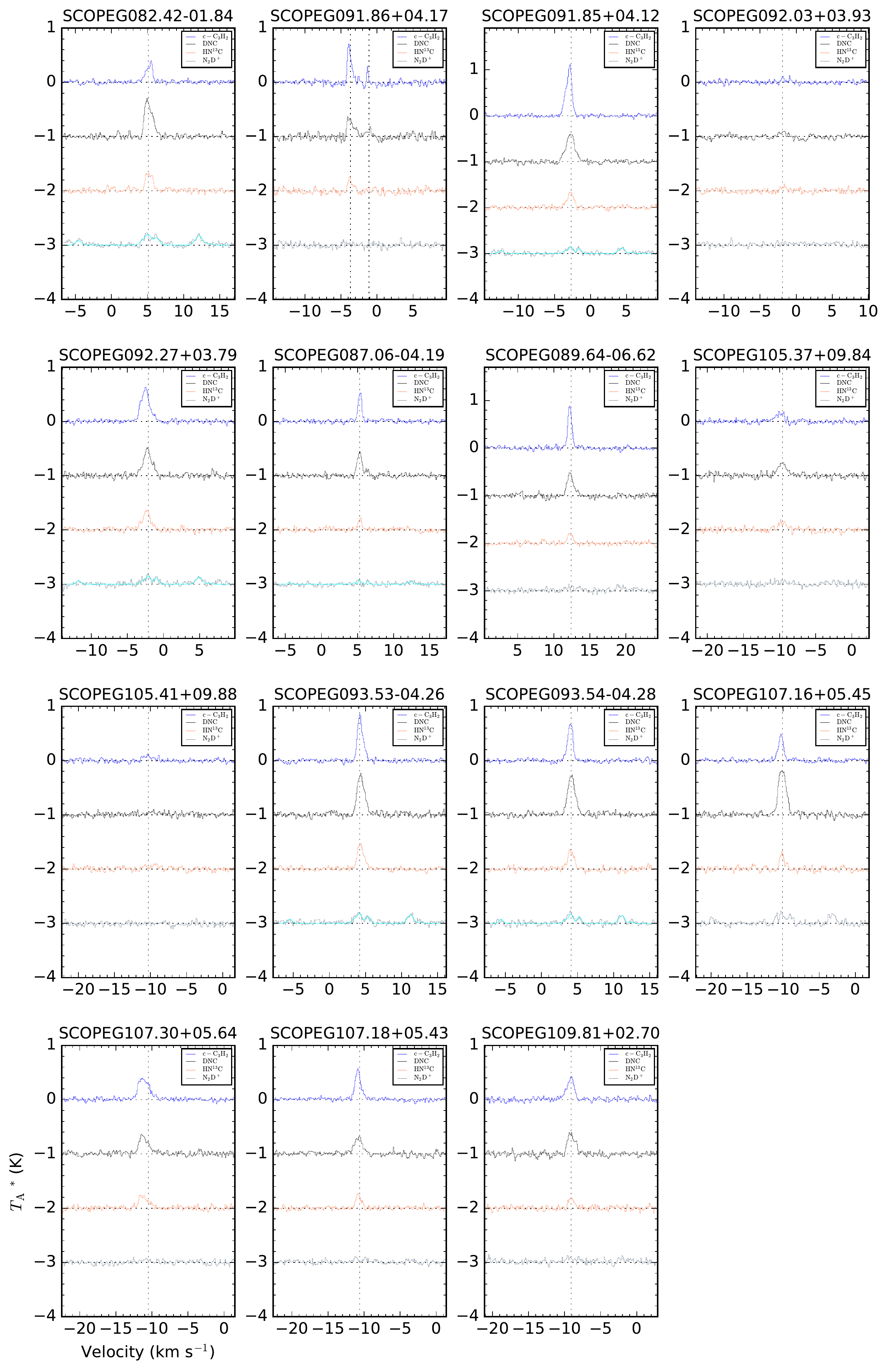}
\caption{(continued)}
\end{figure*}

\allauthors

\end{document}